\documentclass[twocolumn,superscriptaddress,aps,prb,amssymb,amsfonts,amsmath,floatfix]{revtex4-1}
\usepackage{bm}
\usepackage{xcolor}
\usepackage{hyperref}
\usepackage{graphicx}
\usepackage[normalem]{ulem}

\begin{document}

\title{Far-from-equilibrium criticality in the Random Field Ising Model with Eshelby Interactions}%

\author{Saverio Rossi}%
\affiliation{LPTMC, CNRS-UMR 7600, Sorbonne Universit\'e, 4 Place Jussieu, F-75005 Paris, France}
\affiliation{Laboratoire de Physique de l'Ecole Normale Sup\'erieure, ENS, Universit\'e PSL, CNRS, Sorbonne Universit\'e, Universit\'e Paris Cit\'e, F-75005 Paris, France}

\author{Giulio Biroli}%
\affiliation{Laboratoire de Physique de l'Ecole Normale Sup\'erieure, ENS, Universit\'e PSL, CNRS, Sorbonne Universit\'e, Universit\'e Paris Cit\'e, F-75005 Paris, France}

\author{Misaki Ozawa}%
\affiliation{Laboratoire de Physique de l'Ecole Normale Sup\'erieure, ENS, Universit\'e PSL, CNRS, Sorbonne Universit\'e, Universit\'e Paris Cit\'e, F-75005 Paris, France}
\affiliation{Univ. Grenoble Alpes, CNRS, LIPhy, 38000 Grenoble, France}

\author{Gilles Tarjus}%
\affiliation{LPTMC, CNRS-UMR 7600, Sorbonne Universit\'e, 4 Place Jussieu, F-75005 Paris, France}

\date{\today}%

\begin{abstract}
We study a quasi-statically driven random field Ising model (RFIM) at zero temperature with interactions mediated by the long-range anisotropic Eshelby kernel. 
Analogously to amorphous solids at their yielding transition, and differently from ferromagnetic and dipolar RFIMs, the model shows a discontinuous magnetization jump associated with the appearance of a 
band-like structure for weak disorder and a continuous magnetization growth, yet punctuated by avalanches, for strong disorder. Through a finite-size scaling analysis 
in 2 and 3 dimensions we find that the two regimes are separated by a finite-disorder critical point which we characterize. We discuss similarities and differences 
between the present model and models of sheared amorphous solids. 
\end{abstract}

\maketitle

A large variety of phenomena on very different scales involve abrupt collective responses to an applied force or field, known as avalanches. Qualitative 
similarities and observed scale invariance in the distribution of these avalanches have prompted the development of simple models whose relevance to actual 
physical situations has been argued within the framework of universality and renormalization group~\cite{fisher98,sethna2001crackling,alava06,
dahmen-BZ11,sethna-zapperi17,perez-reche16,wiese22}. The systems of interest involve quenched disorder in one form or another, interactions 
between a large number of degrees of freedom, and evolve far from equilibrium. As thermal fluctuations are generally suppressed and driving 
rates very slow, a first conceptual simplification is to consider the limit of a quasi-static driving at zero temperature. 

Scale invariance in such athermal quasi-statically (AQS) driven disordered systems can still appear in quite diverse contexts, which one can sort out according to 
the required number of fine-tuned control parameters. No fine-tuning at all is needed in the case of 
self-organized criticality~\cite{bak87,bak88} and marginal 
stability~\cite{muller-wyart15}. Fine-tuning the magnitude of the driving force corresponds to the broad class of depinning critical transitions, as in elastic manifolds 
in a random environment~\cite{nattermann92,fisher98,wiese22}. Finally, by tuning an additional parameter, the strength of the disorder, one may have a 
critical point separating a regime with a discontinuous, ``snapping'', response associated with an extensive avalanche from a regime with an essentially continuous, 
``popping'', response punctuated by finite avalanches~\cite{sethna2001crackling}. The archetype of such a disorder-controlled criticality is provided by the AQS driven random-field Ising model 
(RFIM)~\cite{sethna93} which is used to describe hysteresis and ``crackling noise'' across a variety of systems~\cite{sethna2001crackling,sethna05}. We focus 
here on this third type of scale invariance and on the associated universality classes. It is well known that 
the dimension of space and the characteristics 
of the order parameter are main factors determining the universality class. Our aim is to understand the role played by nature of the interactions. Note that 
symmetries such as time-reversal and internal symmetries are expected to be less important in these far-from-equilibrium critical points 
because they are anyhow broken by the AQS dynamics.

A broad and important class of AQS driven disordered systems is represented by sheared amorphous solids. When an amorphous material such as glass is 
uniformly deformed from an initial quiescent state, one may observe two different types of yielding behavior: brittle yielding, where the sample 
abruptly breaks with a strong strain localization in the form of a system-spanning shear band, and ductile yielding, in which the sample gradually deforms and flows 
with the presence of subextensive avalanches. 
The nature of the yielding transition depends on the stability and the degree 
of structural and mechanical disorder of the material, which can somewhat be varied through the preparation protocol such as the cooling rate for a glass~\cite{rodney2011modeling,kumar2013critical,vasoya2016notch,fan2017effects,ozawa2018random}. Poorly-annealed samples, which have a lower stability 
and a higher degree of disorder, show ductile yielding, while well-annealed samples with a higher stability and less disorder display brittle yielding. It has been 
argued that the two types of yielding behavior are separated by a disorder (or stability) controlled critical point akin to that of an AQS driven 
RFIM~\cite{ozawa2018random,ozawa2020role}. Although debated~\cite{barlow2020ductile,richard2021finite,pollard2022yielding}, this proposal 
has recently received strong support from our extensive numerical study of a uniformly sheared mesoscopic elasto-plastic model~\cite{rossi2022finite}. 
The similarity between the transition pattern in the RFIM and in mesoscopic models of uniformly sheared amorphous solids has been 
established at the mean-field level~\cite{ozawa2018random,rossi2022emergence,truski20}, but in finite dimensions the geometry and the spatial structure of the large avalanches are clearly 
different in sheared glasses and in the conventional RFIM. 
One major ingredient that is missing in the standard RFIM model is the anisotropy and long-range nature of the elastic interaction (couplings are instead short-range, isotropic and all negative).  
Therefore, in order to mimic the quadrupolar plastic events observed in deformed amorphous materials 
and their interactions  leading to the appearance of shear bands we replace the nearest-neighbor ferromagnetic interactions of the RFIM by the Eshelby propagator 
of linear elasticity~\cite{eshelby1957determination,picard2004elastic}. This leads to an Eshelby-RFIM with anisotropic long-ranged couplings:
\begin{equation}
    \label{eq:Hamiltonian_Eshelby_RFIM}
    \mathcal{H}(\{s_i\}) = - \frac 12\sum_{i,j}G_{ij} s_i s_j - \sum_{i} (h_i+H) s_i,
\end{equation}
where the Ising spin variables $s_i = \pm 1$ are placed at the vertices of a $d$-dimensional cubic lattice, $h_i$ is the local random field, independently 
chosen from a normal distribution of zero mean and variance $R^2$, $H$ is the applied magnetic field, and $G_{ij}\equiv G({\bf r}_{ij})$ is the Eshelby kernel, 
which decays with distance as $r_{ij}^{-d}$ and has the angular dependence of a quadrupole-quadrupole interaction, {\it e.g.}, in $2d$, 
$G({\bf r}_{ij}) = \cos(4 \theta_{ij})/(\pi r_{ij}^2)$, 
where $\theta_{ij}$ is the angle relative to the $x$-axis. This kernel explicitly breaks isotropy (beyond mere lattice effects), the $x$-axis playing the role of the 
direction of shear. To avoid spin self-interaction we moreover fix $G_{ii}=0$. More details on the interaction kernel are given in the Supplemental Material (SM).

Our goal is then two-fold: first, to investigate the issue of universality classes for disorder-controlled criticality in AQS driven disordered systems and, second, 
to explore the possibility of building an effective theory for the yielding transition of uniformly sheared amorphous solids. In this respect, the description provided by elasto-plastic models    (EPMs)~\cite{picard2004elastic,nicolas2018deformation}
falls short because although a drastic phenomenological simplification of the real processes at play, it is still too complex to be analyzed with accurate and controlled theoretical methods. In order to achieve our goal, we perform extensive numerical simulations of the Eshelby-RFIM in two and three dimensions, and we compare the results to their counterparts obtained for EPMs. (Direct numerical simulations of atomistic models are limited to sizes which do not allow a thorough finite-size scaling analysis.)

We follow the AQS dynamics of the model starting from a stable initial condition with all spins pointing down at a large negative field $H=-50$. The field is then 
increased (we study the ascending branch of the hysteresis loop) until a first spin becomes unstable. A spin $s_i$ becomes unstable when its effective 
local field $h_i^{\rm eff} = \sum_{j \neq i}G_{ij} s_j + h_i+H$ 
becomes such that $h_i^{\rm eff} s_i<0$.  This may then lead other spins to become unstable, thereby leading to a collective avalanche which stops when all spins are stable again. Note that since the interaction kernel is not positive definite the dynamics is no longer Abelian and the order in which the spins are flipped in an avalanche may change the precise outcome~\cite{magni1999hysteresis,kurbah11,chakraborty17}. In the present work we choose an update in which at each 
step we pick at random one of the unstable spins to flip it and we recalculate the stability of all other spins (see the SM). To properly average over the random-field disorder and perform finite-size scaling analyses when necessary, we consider systems of linear size $L=N^{1/d}$ from $256$ to $2048$ in $d=2$ and $48$ to 
$128$ in $d=3$ with $200$-$3000$ samples for each size and we use periodic boundary conditions.

\begin{figure}
\centering
\includegraphics[width=0.8\linewidth]{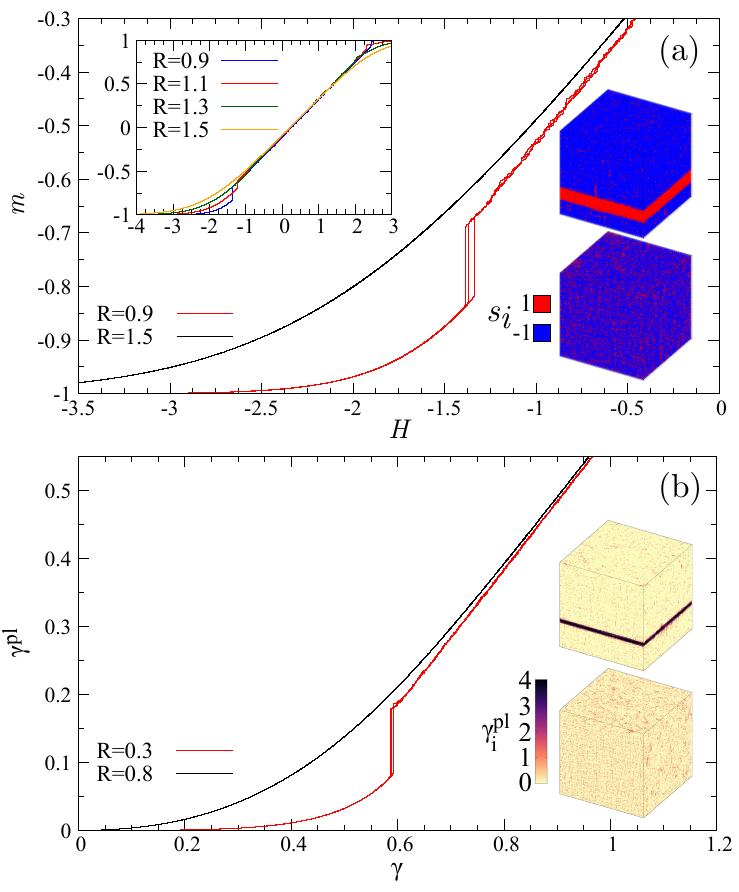}
\caption{(a): Magnetization $m$ versus applied field $H$ in the $3d$ Eshelby-RFIM for $R=1.5$ (black) and $R=0.9$ (red) (3 independent samples with $L=128$). 
Upper left inset: $m(H)$ over a larger range of $H$ up to $m=+1$ (here $L=104$). Lower right inset: Spin configurations at $H=-1.2$ for $R=0.9$ (top) and $R=1.5$ (bottom).
(b): Cumulative volume-averaged plastic strain $\gamma^{\rm pl}$ versus imposed strain $\gamma$ for the $3d$ uniformly sheared EPM studied in Ref.~[\onlinecite{rossi2022finite}] and two 
values of the initial disorder $R$, $R=0.3$ (brittle) and $R=0.8$ (ductile). Inset: Real-space configurations of the local plastic activity $\gamma_i^{\rm pl}$ at $\gamma=0.7$ for $R=0.3$ (top) and $R=0.8$ (bottom).}
\label{fig:largesample_magnetization}
\end{figure}

Our first result is to show that changing the nature of the interaction to an Eshelby kernel indeed drastically modifies the geometry and the spatial structure of the large avalanches. 
We illustrate the transition pattern as a function of the disorder strength $R$ by plotting the volume-averaged magnetization $m=(1/N)\sum_{i=1}^N s_i$ versus $H$ and some 
spin configurations at selected values of $H$ for a few typical $3d$ samples in Fig.~\ref{fig:largesample_magnetization}(a). For a large $R$ the evolution is 
rather smooth and the avalanches are subextensive and uniformly distributed in space. On the other hand for a small $R$, the magnetization displays a jump of O(1) at 
a specific ``coercive field'' $H_{\rm co}$ that is followed by an apparently linear regime. In both cases the magnetization saturates at $+1$ for a large enough 
positive applied field. The striking feature when comparing to the behavior of the convential RFIM~\cite{sethna93,sethna2001crackling,sethna05} is that the jump 
is associated with a system-spanning avalanche that is spatially localized in a band (see the inset). This is reminiscent of the shear band found in the 
brittle yielding of sheared amorphous solids. The similarity can be made even more vivid by looking at the cumulative plastic strain in the AQS evolution of an 
elasto-plastic model (EPM), which is the accumulated plastic activity $\gamma_i^{\rm pl}$ at each site $i$ averaged over the whole lattice. In Fig.~\ref{fig:largesample_magnetization}(b) we plot the results for the $3d$ model that we have studied in Ref.~[\onlinecite{rossi2022finite}] for two values of the initial 
disorder corresponding to ductile and brittle yielding. The correspondence between the two models is that a down spin represents a site that has not yielded and 
an up spin a site that has had a plastic event. Note that the presence of a magnetization  jump corresponding to a single band 
of flipped spins is not found in the previously studied RFIM models, including those with long-range dipole-dipole interactions~\cite{sethna406320random,magni1999hysteresis,dubey15}, 
where one observes either a relatively isotropic macroscopic avalanche~\cite{sethna406320random,sethna2001crackling,vives-perez03,vives-perez04} or exotic labyrinthine patterns~\cite{magni1999hysteresis,dubey15}. This suggests that the Eshelby kernel with its specific quadrupolar symmetry is the key ingredient for reproducing 
the strong band-like behavior~\footnote{It has been argued in Ref.~[\onlinecite{procaccia_screening}] that the long-range decay of
the Eshelby kernel is actually screened by the presence of
emergent dipoles.Work is in progress to test the robustness
of our results to the range of the quadrupole-quadrupole
interaction.}.

We now consider the discontinuous regime at small disorder and the mechanism of band formation. The magnetization can be taken as an order parameter but we 
find it more convenient to also introduce an alternative order parameter which better captures the appearance of a band.  In each sample we monitor $m_{\rm b}$, 
the maximum of all the line-averaged magnetizations in $2d$ and plane-averaged magnetizations in $3d$, where lines/planes are perpendicular to the 
axes: $m_{\rm b}=-1$ at the beginning of the process and it abruptly jumps to a value close to $+1$ at the sample-dependent coercive field $H_{\rm co}$. Averaging 
over disorder, {\it i.e.}, samples, which we indicate by an overline, rounds the discontinuity, but a finite-size scaling of the associated susceptibilities shows that they diverge 
with $L$ at a well-defined coercive field (see the SM), so that the discontinuity persists in the thermodynamic limit.

\begin{figure}
\centering
\includegraphics[width=\linewidth]{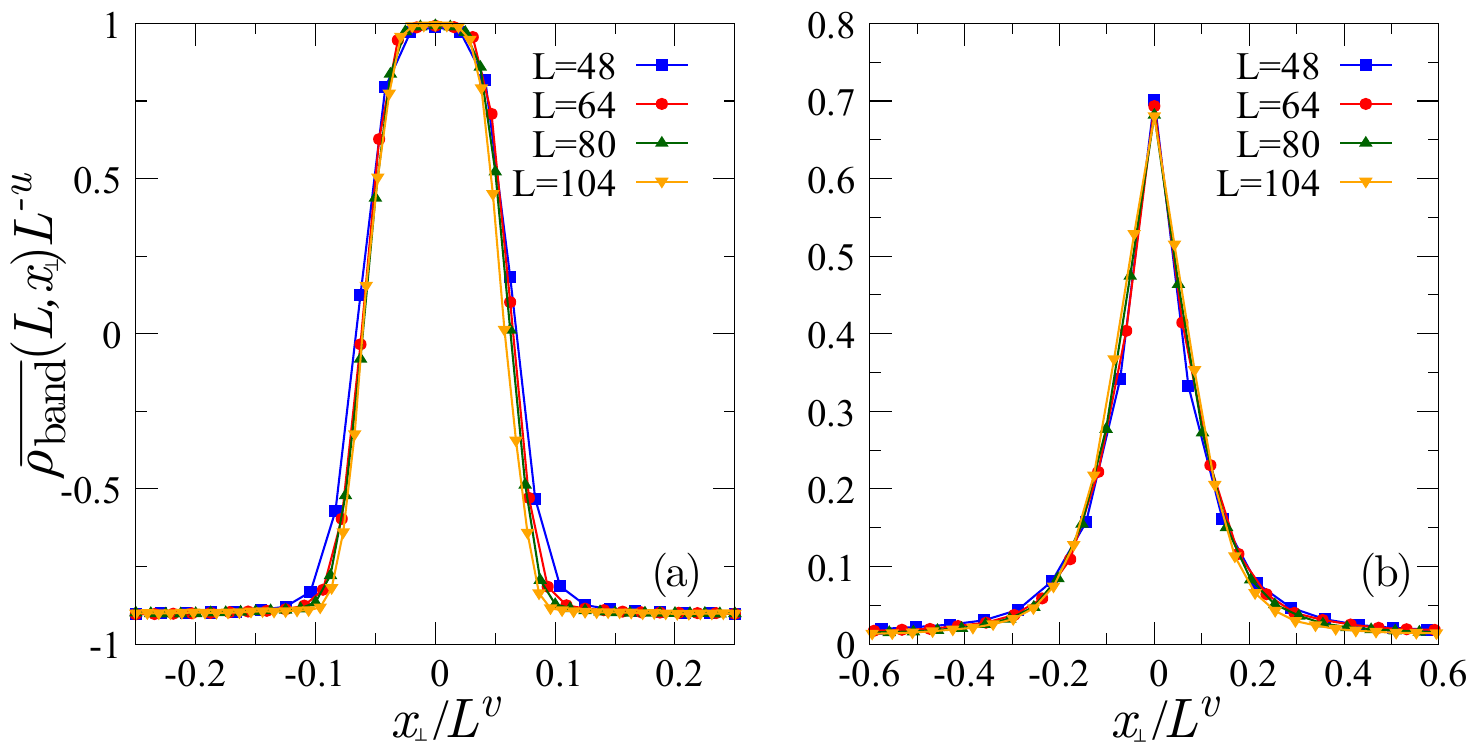}
\caption{Scaling plot of the average band profiles $\overline{\rho_{\rm band}}(L,x_{\perp})/L^u$ versus $x_{\perp}/L^v$ for the $3d$ Eshelby-RFIM (a) and $3d$ EPM (b) right 
at the discontinuous transition, obtained for $R=0.9$ and $R=0.25$, respectively. The collapses are obtained for $u=0$ and $v=1$ for the Eshelby-RFIM, and 
$u= 0.32$ and $v= 0.68$ for the EPM.}
\label{fig:shearband_distribution}
\end{figure}

A major difference between the Eshelby RFIM and the EPMs is the possibility for a site to undergo many repeated events: a spin essentially flips once over the 
whole evolution of the system (see the SM) whereas a site in a uniformly sheared EPM can yield a large number of times. This is particularly important in the 
process of band formation. To contrast the behavior of the two types of models we compute the band profile along the direction $x_{\perp}$ perpendicular to band propagation 
right at the coercive field. The band profile for the Eshelby-RFIM is given by 
$\rho_{\rm band}^{\rm RFIM}(L,x_{\perp}) = (1/L^{d-1}) \sum_{\mathbf{ x}_{\parallel}}s_{(\mathbf{ x}_{\parallel},x_{\perp})}$ and that for the EPM by $\rho_{\rm band}^{\rm EPM}(L,x_{\perp}) = (1/L^{d-1}) \sum_{\mathbf{ x}_{\parallel}}\gamma^{\rm pl}_{(\mathbf{ x}_{\parallel},x_{\perp})}$,
where $(\mathbf{ x}_{\parallel},x_{\perp})\equiv i$ specify the location of a site and we have shifted $x_{\perp}$ in each sample such that $x_{\perp}=0$ corresponds to the center of the band.
In the RFIM, a jump of O(1) in the magnetization implies that the associated band volume scales as $N=L^d$. As 
$\overline{\rho_{\rm band}^{\rm RFIM}}(L,x_{\perp}=0)$ should be essentially independent of $L$, one then expects that the band width scales as $L^v$ with $v=1$, 
so that the change of magnetization summed over the whole band goes as $L^d$. This is verified by the finite-size scaling result illustrated in 
Fig.~\ref{fig:shearband_distribution}(a) for the $3d$ case. On the other hand, the number of plastic events per site in the shear band of the EPM appears to 
grow with the system size. This translates into $\overline{\rho_{\rm band}^{\rm EPM}}(L,x_{\perp}\approx0)\sim L^u$ with $u>0$ while, concomitantly, the width of the 
shear band scales as $L^v$ with $v=1-u<1$. This leads to a difference between the scaling of the number of sites and of plastic events: the total number of {\it sites} in the shear band goes as $L^{d-1+v}$ while the total number of 
{\it plastic events} in the shear band goes as $N=L^d$. 
We find that $v\approx 0.68$ and $u\approx 0.32$ for the $3d$ EPM: see Fig.~\ref{fig:shearband_distribution}(b). In the SM we also provide additional evidence of the difference between Eshelby-RFIM and EPM  in the weak-disorder, discontinuous regime by looking at the avalanche distribution and the marginal stability 
property.


\begin{figure}
\centering
\includegraphics[width=\linewidth]{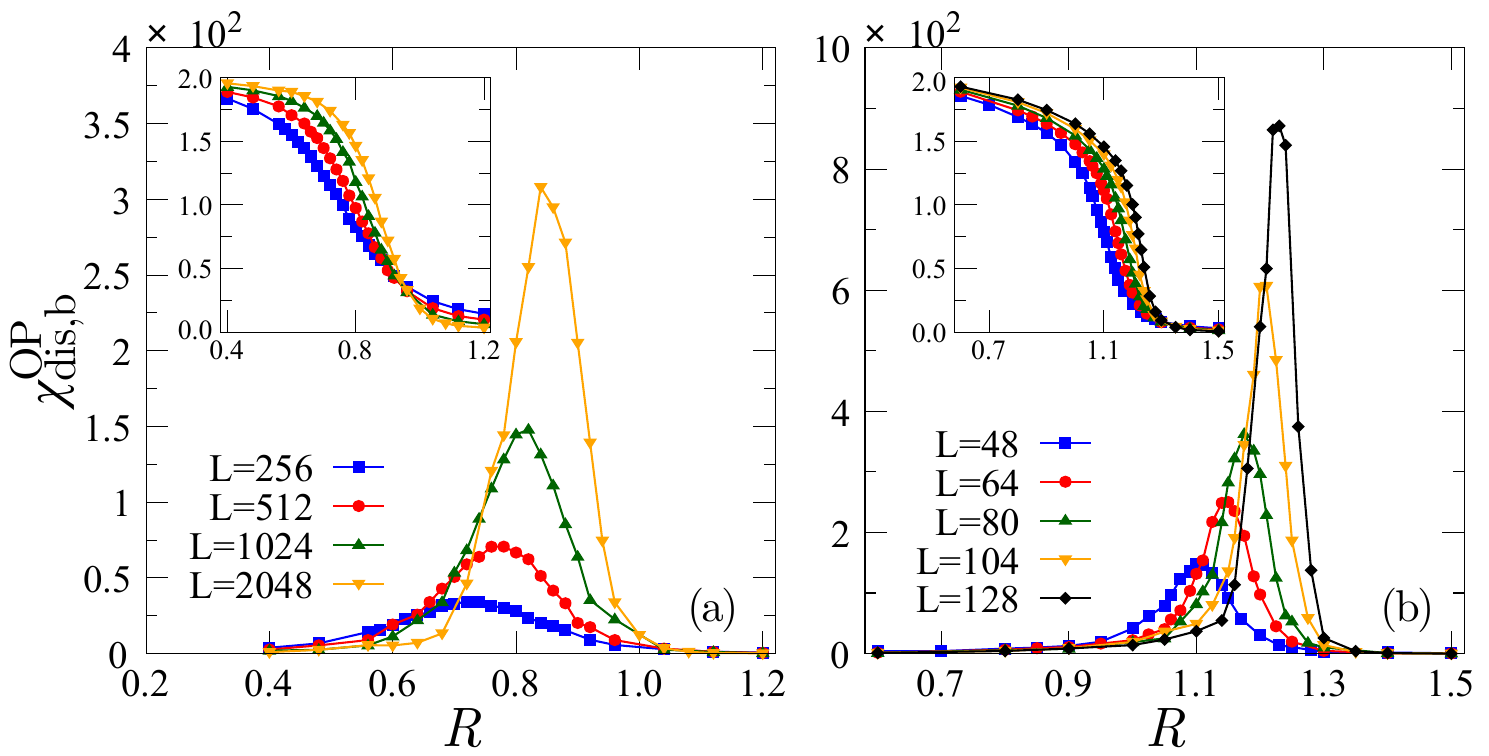}
\caption{Disconnected susceptibility $\chi^{\rm OP}_{\rm dis,b}(R,L)$ and associated disorder-averaged order parameter $\overline{\Delta m_{\rm b,max}}$ 
(insets) versus disorder strength $R$ for several linear system sizes $L$ for the $2d$ (a) and the $3d$ (b) Eshelby-RFIM.}
\label{fig:critical_point}
\end{figure}

Between the discontinuous regime at small $R$ and the crossover one at large $R$ one expects a disorder-controlled critical point. To pinpoint the latter 
and characterize its properties we have performed a finite-size scaling analysis. Inspired by our previous EPM work~\cite{rossi2022finite} we have  
introduced two different order parameters, the maximum jump of the magnetization, $\Delta m_{\rm max}$, and that of the quantity $m_{\rm b}(H)$ (see above), 
$\Delta m_{\rm b, max}$, where the maximum is taken for each sample over a whole AQS evolution from $H=-50$ to a very large $H$ (until all spins become 
positive). By construction, both are nonzero of O(1) when the sample displays a strong band-like avalanche with a discontinuous magnetization jump, whereas they 
are nearly $0$ for a gradual evolution of the magnetization curve. As an illustration we display in Fig.~\ref{fig:critical_point} the disorder-averaged 
value $\overline{\Delta m_{\rm b, max}}(R,L)$ and the associated disconnected susceptibility $\chi_{\rm dis,b}^{\rm OP}(R,L)=L^{d-1}{\rm Var}(\Delta m_{\rm b, max})$, where Var denotes the variance, in $2d$ and $3d$. The behavior observed as a function of the linear system size $L$ is typical of what is expected for a critical point at a value $R_c(L)$ corresponding to the position of the susceptibility peak. Similar curves (but with more noise) are obtained for $\overline{\Delta m_{\rm max}}(R,L)$ and the associated disconnected susceptibility 
$\chi_{\rm dis}^{\rm OP}(R,L)=L^{d}{\rm Var}(\Delta m_{\rm max})$. We analyze the results through a finite-size scaling ansatz in which we assume that, despite 
the anisotropy of the interactions, criticality is characterized by a unique diverging correlation length with exponent $\nu$~\footnote{Although hard to prove directly, a first hint in this direction is provided by the behavior of the system-spanning band in the discontinuous regime: it is clearly anisotropic but, as shown above, its size nonetheless scales linearly with $L$ both in the longitudinal and the transverse direction.
}:
\begin{equation}
\begin{aligned}
\label{eq:FSS}
&\overline{\Delta m_{\rm max}} \sim L^{-\frac{\beta}{\nu}} \mathcal M(r L^{\frac 1{\nu}}), \,
\overline{\Delta m_{\rm b, max}} \sim L^{-\frac{\beta_{\rm b}}{\nu}} \mathcal M_{\rm b}(r L^{\frac 1{\nu}}),
 \\&
\chi_{\rm dis}^{\rm OP}(R,L)\sim L^{\frac{\bar\gamma}{\nu}}\,\bar\Psi(r L^{\frac 1{\nu}}),\,
\chi_{\rm dis, b}^{\rm OP}(R,L)\sim L^{\frac{\bar\gamma_{\rm b}}{\nu}}\,\bar\Psi_{\rm b}(r L^{\frac 1{\nu}}),
\end{aligned}
\end{equation}
where $r=(R-R_c(L))/R$ is the relative distance from the critical disorder, $\beta$, $\beta_{\rm b}$, $\bar\gamma$, $\bar\gamma_{\rm b}$ are critical exponents, 
and $\mathcal M$, $\bar\Psi$, $\mathcal M_{\rm b}$, $\bar\Psi_{\rm b}$ are scaling functions. Note that by construction,  $\bar\gamma/\nu \leq d$, 
$\bar\gamma_{\rm b}/\nu \leq d-1$ and, for consistency,  $\beta/\nu , \,\beta_{\rm b}/\nu \geq 0$.

\begin{figure}
\centering
\includegraphics[width=\linewidth]{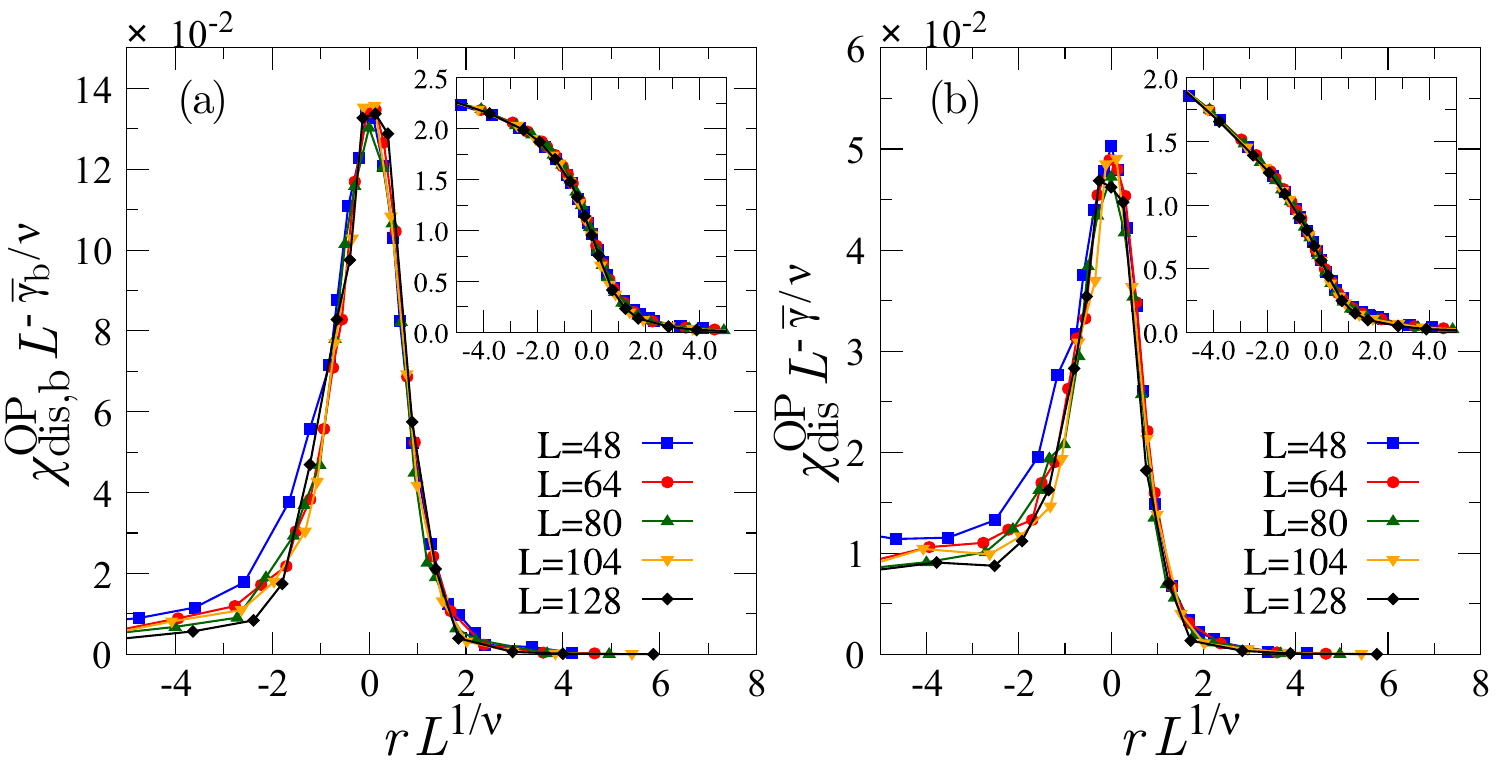}
\caption{Finite-size scaling collapse of the disconnected susceptibilities $\chi^{\rm OP}_{\rm dis,b}(R,L)$  with $\bar\gamma_{\rm b} / \nu= 1.81$ (a) and 
$\chi^{\rm OP}_{\rm dis}(R,L)$ with $\bar\gamma/\nu=1.7$ (b) in $3d$. 
Insets: Same for $\overline{\Delta m_{\rm b,max}}$ with $\beta_{\rm b}/\nu=0.08$ (a) and $\overline{\Delta m_{\rm max}}$ with 
$\beta/\nu=0.62$ (b). In all cases $\nu=1.4$. Compare with Eq.~(\ref{eq:FSS}).}
\label{fig:collapse}
\end{figure}

We have used four different methods to determine the critical exponents and functions: i) power-law fits to the maximum and the width of the susceptibilities, ii) power-law fit 
to the drift of the critical disorder, $R_c(\infty)-R_c(L)\sim L^{-1/\nu}$, iii) empirical scaling collapses of the data according to Eq.~(\ref{eq:FSS}) 
by adjusting the exponents, and iv) a fit to a (flexible) analytical form for the susceptibility master-curves~\cite{sethna2013percolationfailure}. We have also checked that compatible (but not as good) results 
for the exponents are obtained by looking at the disconnected susceptibilities defined from $m(H)$ and $m_{\rm b}(H)$ (instead of $\Delta m_{\rm max}$ and 
$\Delta m_{\rm b, max}$) when evaluated at their maximum over $H$ and at the critical disorder $R_c(L)$. On the other hand, the connected susceptibilities which 
must be obtained as numerical derivatives with respect to $H$ cannot be reliably computed and in any case vary much less with $L$ than the disconnected ones. 
Details are provided in the SM. In Fig.~\ref{fig:collapse} we illustrate the outcome with a collapse of the disconnected susceptibilities and the corresponding average order parameters (insets) for the $3d$ case.
We find for the $3d$ Eshelby-RFIM
\begin{equation}
\begin{aligned}
\label{eq:3Dexponents}
&\bar\gamma/\nu\approx 1.65-1.75, \; \beta/\nu\approx 0.58-0.66, \; \nu\approx 1.2-2.4, \\&
\bar\gamma_{\rm b}/\nu\approx 1.77-1.9, \; \beta_{\rm b}/\nu\approx 0.04-0.12, 
\end{aligned}
\end{equation}
where the rather large error bars include the different methods of determination mentioned above.  (We note that the procedures are less sensitive to the value of  $1/\nu$ than 
to that of the other exponents, so that $\nu$ is more poorly determined.) The exponents satisfy the bounds given above and one can also push further the 
scaling ansatz to relate them. Indeed, at criticality one expects that the spatial correlations of the sample-to-sample spin fluctuations become scale-free with a power-law exponent related to that of the divergence of the disconnected susceptibility $\chi_{\rm dis}^{\rm OP}$. If one assumes that the anisotropy manifests itself in the prefactor but not in the power-law dependence, this exponent is simply $d-\bar\gamma/\nu$ (an integral over the volume then gives back the divergence of $\chi_{\rm dis}^{\rm OP}$). The scaling dimension of the average order parameter at the critical point is then $-(d-\bar\gamma/\nu)/2$. 
A similar reasoning holds for the quantities defined from $\Delta m_{\rm b, max}$, except that the relevant correlation function is now for spins within a line 
(in $2d$) or a plane (in $3d$) associated with the incipient critical band. This new correlation function is expected to decay as a power law at large distance 
with exponent $d-1-\bar\gamma_{\rm b}/\nu$  (an integral over the line or the plane then gives back the divergence of $\chi_{\rm dis, b}^{\rm OP}$), so that the 
average order parameter should scale with system size with an exponent $-(d-1-\bar\gamma_{\rm b}/\nu)/2$. Putting this together leads to
$2 \beta/\nu= d-\bar\gamma/\nu$ and $2 \beta_{\rm b}/\nu= d-1-\bar\gamma_{\rm b}/\nu$, 
which is verified within error bars by the values in Eq.~(\ref{eq:3Dexponents}). These results provide a strong support for the presence of a critical point at 
a finite value of the disorder strength in $3d$. Note that the values in Eq.~(\ref{eq:3Dexponents}) seem to indicate that 
$\bar\gamma/\nu \approx \bar\gamma_{\rm b}/\nu$ and $\beta/\nu \approx \beta_{\rm b}/\nu+(1/2)$, but we have not been able to find strong arguments to 
support this finding beyond invoking Occam's razor.

The situation in $2d$ is more problematic as we find that $\bar\gamma_{\rm b}/\nu \approx 1.00-1.05$ (but being larger than $1$ is forbidden by the bound) and 
$\beta_{\rm b}/\nu \approx -0.04-0.00$ (but being negative is not physical). Correspondingly, we roughly estimate that $\bar\gamma/\nu \approx 1.05-1.25$, 
$\beta/\nu \approx 0.45-0.48$, and $\nu \approx 2.3-3.0$ (see the SM). The model does not seem to be at its lower critical dimension because, 
contrary to what would then be expected, the critical disorder $R_c(L)$ slowly {\it increases} with $L$ (see Fig.~\ref{fig:critical_point}) instead of decreasing as, 
{\it e.g.}, found  in the standard AQS driven RFIM in $d=2$~\cite{spasojevic2011numerical,hayden2019unusual}.  This points to a discontinuous transition 
for $m_{\rm b}$, {\it i.e.}, the sudden emergence of a positively magnetized {\it line} at some $R_c$, but the band width itself appears to scale sub-linearly 
with $L$, like $L^{1-\beta/\nu}$. This corresponds to an unusual scenario in which the disorder-controlled transition between the discontinuous and the 
continuous regimes is a fluctuation-induced discontinuous one for the band-order parameter but a continuous one for the magnetization.

We can finally compare the critical behavior of the Eshely-RFIM just obtained with that of the EPM studied in Ref.~[\onlinecite{rossi2022finite}]. As already mentioned, 
we find that the qualitative features (transition pattern, shape of the large-scale avalanches, dominance of the sample-to-sample fluctuations) are very similar. We 
also observe that the critical exponents in the two models are close and evolve in the same direction between $3d$ and $2d$.  The better-determined ratio 
$\bar\gamma_{\rm b}/\nu$ is about $10\%$ higher in the Eshelby-RFIM than in the EPM~\footnote{Note that the EPM exponent was defined differently in Ref.~[\onlinecite{rossi2022finite}] and shifted by $+1$ because of a different definition of $\chi_{\rm dis,b}^{\rm OP}$.} but, considering the difficulty to reliably extract exponent values in 
out-of-equilibrium disordered systems~\cite{perkovic1996,*perkovic1999}, this difference may not be significant enough for reaching a definite conclusion. 
Furthermore, the exponents $\bar\gamma/\nu$ and $\beta/\nu$ are compatible between the two models within error bars and, as displayed in Fig. S8 of the SM, 
the disconnected susceptibilities $\chi_{\rm dis}^{\rm OP}(R,L)$ in the $3d$ case, where they can be more reliably computed, all collapse very well onto the same 
master-curve. This suggests that the Eshelby-RFIM and the EPM are in the same universality class. However, even larger system sizes with then still more 
sophisticated scaling analyses ({\it e.g.}, a detailed characterization of the system-spanning avalanches at criticality as in Refs.~[\onlinecite{vives-perez03,vives-perez04}] or of the spatial 
correlation functions) will be necessary to definitely settle the issue. On the other hand, we have shown ample evidence that the Eshelby-RFIM and the ferromagnetic 
RFIM clearly belong to distinct universality classes.

To sum up, we have introduced and numerically studied a version of the athermally and quasi-statically driven RFIM in which the interactions have the anisotropic and long-ranged form of the Eshelby propagator used in the theory of amorphous solids under deformation. We have found that the model shows a disorder-controlled 
transition between a discontinuous and a continuous regime, as the standard RFIM but with a different phenomenology that is instead very similar to that found 
in elasto-plastic models (EPMs) describing the yielding transition of uniformly sheared amorphous materials. Away from the 
disorder-controlled transition, all of these models show quite distinct behavior. This is seen either in the form of the macroscopic avalanches or in the number of times 
a site is active during the full out-of-equilibrium evolution of the system. Universality, if present, holds at the disorder-controlled critical point only, and this is what 
would justify the search for simplified models, such as the Eshelby-RFIM just introduced, as an effective theory to describe the yielding transition of amorphous solids. 
The transition is indeed critical in $3d$ (and is unusual with a mixed character in $2d$). Many observations suggest that Eshelby RFIM and EPMs 
are in the same universality class, although a definite conclusion cannot be reached at this point. In addition, our results clearly show that the Eshelby and the 
ferromagnetic RFIM belong to different universality classes. They are also qualitatively different than those found in RFIMs with dipole-dipole interactions. For future 
work, an exhaustive study of the effect of the interactions on the transition pattern would be extremely valuable, in particular to disentangle the role of the anisotropy 
and that of the range of the interaction.

{\it Acknowledgments:}
We thank James Sethna and Francesco Zamponi for fruitful discussions. We also thank J. Sethna for suggesting an efficient way to implement the Eshelby kernel.
This work was granted access to the HPC resources of the MeSU platform at Sorbonne Universit\'e and was supported by the Simons Foundation Grant 
No. 454935 (G.B.).

\bibliography{references}

\clearpage

\section*{Supplemental Material}
\subsection{AQS driven Eshelby random-field Ising model}
\subsubsection{Model and AQS evolution}
\label{app_sub:model}

We study two and three-dimensional lattice-based random-field Ising models (RFIMs) with periodic boundary conditions. The Hamiltonian of the system is given 
by
\begin{equation}
\label{eq:Hamiltonian_SM}
\mathcal{H}(\{s_i\}) = - \frac 12\sum_{i,j} G_{i j} s_i s_j - \sum_{i} (h_i+H) s_i,
\end{equation}
where Ising spins, $s_{i}=\pm 1$, are placed on each of the $N$ sites of the (square or cubic) lattice, local random fields $h_i$ are identically and independently 
taken from the Gaussian distribution $\mathcal{N}(0,R^2)$ of zero mean and standard deviation $R$; $R$ therefore controls the disorder strength. $H$ is the 
applied magnetic field, which is quasi-statically ramped up. The volume-averaged magnetization, $m$, is defined by $m=(1/N)\sum_i s_i$, where $N=L^d$ with 
$L$ the linear box length and $d$ is the spatial dimension.

The stability of spin $s_i$ is governed by the effective local field $h_{i}^{\rm eff}$ which is defined as
\begin{equation}
    h_{i}^{\rm eff}= \sum_{j \neq i} G_{ij}s_{j} + h_i + H.
    \label{eq:effective_field}
\end{equation}
When $h_{i}^{\rm eff} s_i >0$, the site $i$ is stable. Otherwise, the spin $s_i$ is unstable and it flips as $s_i \rightarrow -s_i$.

For the initial condition, we start with $s_{i} = -1$ and $h_{i}^{\rm eff}<0$ for all $i$, with $H = -50$. We then increase the external field $H$ until a spin becomes 
unstable, which defines discrete time steps that are labelled by an index $t$. For each time step $t$, the external field is increased by the amount, 
$\Delta H_t=\min_i \{ -h_{i, t}^{\rm eff} \ | \ h_{i, t}^{\rm eff}<0 \}$, which is the minimum field increment to change the sign of $h_{i^*, t}^{\rm eff}$ (and hence 
the sign of $s_{i^*, t}$) at the site $i^*$ that is closest to instability. 

When a spin  $s_i$ flips, it influences the local effective field of all the other spins $j \neq i$ as
\begin{equation}
h_{j,t}^{\rm eff} \rightarrow h_{j,t}^{\rm eff} + 2 G_{j i}.
\label{eq:update_effective_field}
\end{equation}
Note that we impose $G_{ii}=0$ to obtain Eq.~(\ref{eq:update_effective_field}), so that the spin flip does not affect its own effective field. This absence of  
self-interaction term is at variance with elastoplastic models (EPMs) where local stress drops take place~\cite{nicolas2018deformation}.
According to Eq.~(\ref{eq:update_effective_field}), the effect of the spin flip at $i$ may trigger an instability for other spins and lead to an avalanche of 
spin flips until all spins become stable again. Then, we increase the external field by the amount $\Delta H_{t+1}$ so that a single spin flips, and the 
process is repeated for this step $t+1$. In this driving mechanism the timescale of the variation of the applied field and that of the development 
of an avalanche are decoupled, which corresponds to the so-called athermal quasi-static (AQS) driving~\cite{maloney2006amorphous}.

To summarize, the time evolution of the model is given by the following algorithm:
\begin{enumerate}
    \item Initialize the spins $s_{i, t=0}=-1$ for all $i$ with $H=-50$ and the local random fields $h_i$ that are i.i.d. from $\mathcal{N}(0,R^2)$;
    \item Find site $i$ that has the largest negative effective field, {\it i.e.}, minimizes $-h_{i,t}^{\rm{eff}}$;
    \item Increase the external field that acts on all sites by $\Delta H_{t} = -h_{i,t}^{\rm{eff}}$ such that only the spin $i$ flips;
    \item Change the effective field acting on each spin following $h_{j,t}^{\rm{eff}} \rightarrow h_{j,t}^{\rm{eff}} + 2 G_{ji}$ (with $G_{ii}=0$);
    \item Check which spins are unstable, {\it i.e.}, do not satisfy $h_{j,t}^{\rm{eff}} s_{j, t} > 0$, choose randomly (see below) one spin among them,  
    and flip it. Propagate the effect of the spin flip and repeat until all sites become stable; 
    \item Repeat 2 - 5.
\end{enumerate}
After each increment of the external field, the local effective field evolves as
\begin{equation}
\label{eq:local_eff_field_evolution}
    h_{i,t+1}^{\rm{eff}} = h_{i,t}^{\rm{eff}} + \Delta H_{t} + \sum_j G_{ij} (s_{j,t+1}-s_{j,t}).
\end{equation}
We then consider the time evolution of the average effective field, $h_{t}^{\rm eff}=(1/N) \sum_i h_{t, i}^{\rm eff}$, which from Eq.~\eqref{eq:local_eff_field_evolution} 
is governed by
\begin{equation}
h_{t+1}^{\rm eff} = h_{t}^{\rm eff} + \Delta H_t + \hat{G}_{{\bf q}={\bf 0}} \Delta m_t,
\end{equation}
where $\hat G_{\textbf{q}={\bf 0}}=\sum_i G_{ij}$,
$\textbf{q}=(q_x,q_y)$ in $2d$ and $\textbf{q}=(q_x,q_y, q_z)$ in $3d$ denote the wave-vector, and $\Delta m_t = m_{t+1}-m_{t}$. 
In this study, we set $\hat{G}_{\mathbf{q}=\mathbf{0}}=0$, so that on average the magnetization change does not influence the change of the effective field.

\subsubsection{Random updating scheme} 

When a spin flips, several other spins can become unstable as a result of the interaction: see Eq.~(\ref{eq:update_effective_field}). This is analogous 
to what happens in elastoplastic models (EPMs) when a site yields locally and in the standard (ferromagnetic) RFIM~\cite{sethna93}. However, contrary to 
the latter whose dynamics has an Abelian property that ensures that the final stable configuration at the end of the avalanche is independent of the order 
in which spins are flipped~\cite{sethna93}, the nonpositive nature of the Eshelby kernel, which leads to ferromagnetic or anti-ferromagnetic interactions depending 
on the vector orientation between sites, breaks this property and the final configuration {\it a priori} depends on the order in which the spins are flipped. 

In our previous investigation of an EPM~\cite{rossi2022finite} we used a parallel update of all unstable spins by means of the Fast Fourier transform algorithm. 
In the case of the Eshelby-RFIM, however, this parallel updating scheme may enter a never-ending loop during simulation. To illustrate this problem, let us 
consider that two spins, $s_i$ and $s_j$, with $G_{ij}<0$ are the only ones unstable at the step $l$ during an avalanche specified by the time $t$. In particular,  
we consider $h^{\rm eff}_{i,t,l} > 0$ and $s_{i,t,l} = -1$ for site $i$, and $h^{\rm eff}_{j,t,l} > 0$ and $s_{j,t,l} = -1$ for site $j$. When $|h^{\rm eff}_{i,t,l}|<2|G_{ij}|$ 
and $|h^{\rm eff}_{j,t,l}|<2|G_{ji}|$, at the next step $l+1$ we obtain
\begin{equation}
    \begin{aligned}
    &h^{\rm eff}_{i,t,l+1} = h^{\rm eff}_{i,t,l} + 2G_{ij} < 0 \quad \textrm{and} \quad  s_{i,t,l+1} = 1, \\    
        &h^{\rm eff}_{j,t,l+1} = h^{\rm eff}_{j,t,l} + 2G_{ji} < 0 \quad \textrm{and} \quad s_{j,t,l+1} = 1, 
    \end{aligned}
\end{equation}
which means that both spins become unstable again at $l+1$. Therefore the process will never stop, making an infinite loop.
We have shown the above example with only a pair of spins for simplicity, but there are many other possible combinations that will lead to a similar situation 
in which the system is stuck in an infinite loop and cannot relax.

In order to avoid this problem, we consider a random updating scheme: once the list of all the unstable sites is established, we choose one site at random 
and flip it. The problem that we had before can be solved as follows. Due to the single updating and stochasticity, we would get
\begin{equation}
    \begin{aligned}
    &h^{\rm eff}_{i,t,l+1} = h^{\rm eff}_{i,t,l} > 0 \quad \textrm{and} \quad s_{i,t,l+1} = 1, \\    
        &h^{\rm eff}_{j,t,l+1} = h^{\rm eff}_{j,t,l} + 2G_{ji} < 0 \quad \textrm{and} \quad s_{j,t,l+1} = -1, 
    \end{aligned}
\end{equation}
which means that both spins are now stable. In practice, we have confirmed that the random updating scheme always converges well, and we have never found 
an infinite loop in simulations.

\subsubsection{Implementation of the Eshelby kernel} 

We obtain the kernel $G_{ij}$ on the lattice by the inverse Fourier transform of $\hat G_{\bf q}$. This is done only once at the beginning of the 
simulation, and the computed $G_{ij}$ is used as a Green's function during the simulation. We first start from a continuum version, $\hat G({\bf q})$ derived 
from linear elasticity theory (see, {\it e.g.}, Ref.~[\onlinecite{picard2004elastic}]): 
\begin{equation}
\label{eqn:four_kernel}
    \begin{split}
  \hat{G}(\textbf{q})\equiv   \hat G(q^2_x,q^2_y) &= - \frac{4 q_x^2 q_y^2}{\left( q_x^2 + q_y^2 \right)^2} \qquad (2d), \\
 \hat{G}(\textbf{q})\equiv    \hat G(q^2_x,q^2_y,q^2_z) &= -\frac{4q_x^2 q_y^2}{q^4} -\frac{q_z^2}{q^2} \qquad (3d).
\end{split}
\end{equation}
Since we study systems of finite size $L$, we discretize the Fourier space.  
Following the procedure described in Ref.~[\onlinecite{ferrero2019criticality}], the wave-vector components are written as 
$q_\mu = (2\pi/L) n_\mu$, with $\mu=x,y,$ and possibly $z$ and $n_\mu=-L/2+1,\cdots,L/2$. Furthermore, we consider a discrete Laplacian adapted 
to the lattice instead of the continuous one, which amounts to replacing $q^2_\mu$ by  $2 - 2\cos q_\mu$ in Eq.~\eqref{eqn:four_kernel}. Notice that both 
in real and Fourier spaces, the kernel is not defined at the origin. Instead we impose the conditions that are discussed above and in the main text, 
namely, $\hat{G}_{\textbf{q}=\textbf{0}}=0$ and $G_{ii}=0$. 
In practice, we use
\begin{equation}
     \hat{G}_{\textbf{q}} = \begin{cases}
     0 \  & (\textbf{q}={\bf 0}) \ ,\\
     \frac{\hat G(q^2_\mu \to 2 - 2\cos q_\mu)}{\mathcal{I}} +\left(1+\frac{1}{L^d-1} \right) \  & (\textbf{q} \neq{\bf 0}) \ ,
     \end{cases}
\end{equation}
where $\mathcal{I}$ is a normalization factor which imposes that $G_{ii}=0$.

After this treatment, the resulting real-space expression of $G_{ij}$ is found to slightly deviate from the $1/r^d$ dependence at short distance (it is smaller). 
{\it Per se} this is not fundamental as the Eshelby propagator is meant to describe the elastic interaction in the far field and does not properly describe what 
happens at short distance. In EPMs, the issue is inessential and only possibly affects the visual aspect of avalanches~\cite{ZapperiBudrikis2013}. In the case 
of the Eshelby-RFIM, we have found that having a too small coupling at short distance prevents the formation of a clear band in $2d$. We have therefore 
manually corrected the nearest-neighbor interaction to generate a $1/r^d$ dependence even at short distance.
\\

\subsection{Order parameters and finite-size scaling analyses}
\label{app_sub:OP}

\begin{figure}
    \centering
 \includegraphics[width=\linewidth]{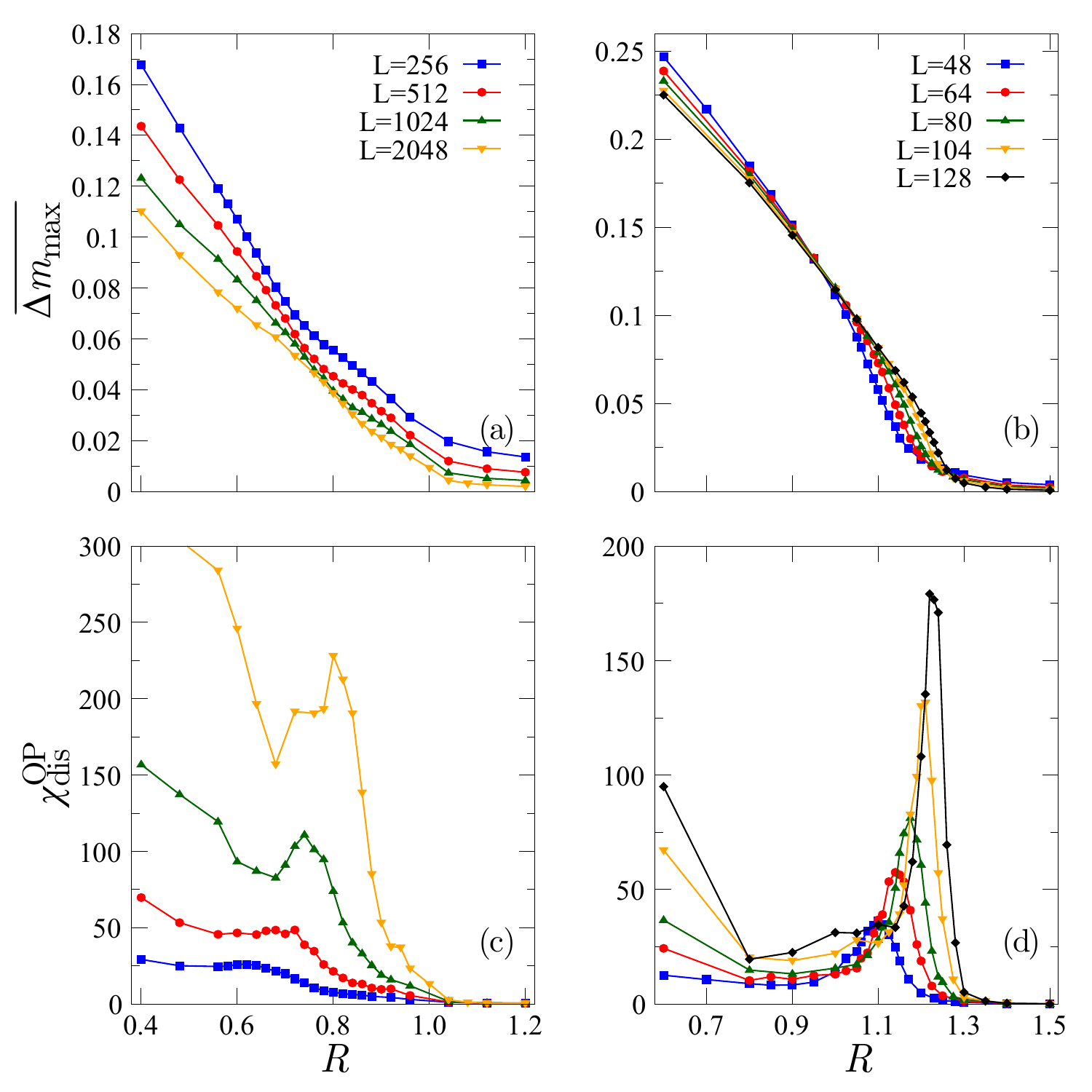}
\caption{Disorder-averaged value of the order parameter, $\overline{\Delta m_{\rm max}}$, as a function of $R$ for several system sizes $L$ in $2d$ (a) and 
$3d$ (b). The corresponding variance of $\Delta m_{\rm max}$ multiplied by $N=L^d$ is the disconnected susceptibility and is shown in $2d$ (c) and $3d$ (d).}
\label{fig:critical_point_delta_m}
\end{figure}

\subsubsection{Order parameters}

The disorder-controlled transition from a discontinuous behavior of the magnetization curve to a continuous behavior can be characterized by several choices of 
order parameter. The most obvious one is the maximum magnetization jump, $\Delta m_{\rm max} = \max_{t}\{\Delta m_t\}$, where $\Delta m_t = m_{t+1} - m_{t}$. 
This is the  analog of the maximum stress drop $\Delta \sigma_{\rm max}$ used in Refs.~[\onlinecite{ozawa2018random,ozawa2020role,bhaumik2021role}]. 
We compute its disorder-averaged value, $\overline{\Delta m_{\rm max}}$, and its variance, 
${\rm Var}(\Delta m_{\rm max})=\left( \overline{\Delta m_{\rm max}^2} - \overline{\Delta m_{\rm max}}^2 \right)$, from which we define the disconnected 
susceptibility as $\chi^{\rm OP}_{\rm dis}(R,L)=L^d {\rm Var}(\Delta m_{\rm max})$. As in the main text, the overline denotes an average over samples, {\it i.e.}, 
realizations of the random fields. The outcome is displayed in Fig.~\ref{fig:critical_point_delta_m} for $2d$ and $3d$ and we see that the disconnected 
susceptibility  shows a peak that grows with the system size at the onset of the growth of $\overline{\Delta m_{\rm max}}$, which is a  hallmark of a critical 
transition. 

However, as we decrease $R$ further, the variance starts to increase again with the system size. This unexpected behavior can be explained by the 
following argument. When $R \to 0$, there are no spin flips as $H$ increases until the discontinuous jump at the coercive field $H_{\rm co}$.
At $H_{\rm co}$, a single site $i^*$ with the largest random field initiates the macroscopic jump. We thus define the maximum random field by 
$h_{\rm max}=h_{i^*}=\max_i\{ h_i \}$. According to Eq.~(\ref{eq:effective_field}), when $R \to 0$, $H_{\rm co}$ and $h_{\rm max}$ are related by 
\begin{equation}
0 = h_{i^*}^{\rm eff} = - \sum_{j \neq i^*} G_{i^*j} + h_{i^*} + H_{\rm co},
\end{equation}
where we have used that $s_{j \neq i^*}=-1$. Since $\sum_i G_{ij}= \hat G_{{\bf q}={\bf 0}}=0$ and $G_{ii}=0$ by definition, we arrive at 
$H_{\rm co}=-h_{\rm max}$. As also described in the main text, the magnetization $m$ grows linearly with the external field $H$ beyond $H_{\rm co}$, 
due to the propagation of the band in the transverse direction. This allows us to write the evolution of $m$ when $R \to 0$ as
\begin{equation}
     m(H) = \begin{cases}
     -1 \ ,  & H < H_{\rm co} , \\
     AH + B  , & H \geq H_{\rm co} ,
     \end{cases}
\end{equation}
where $A$ and $B$ are some constants. Thus, we obtain $\Delta m_{\rm max} = m(H_{\rm co}) + 1 = A H_{\rm co} + B + 1$. 
We can then relate the variance of $\Delta m_{\rm max}$, $H_{\rm co}$, and $h_{\rm max}$ through
\begin{eqnarray}
{\rm Var}(\Delta m_{\rm max}) = A^2\, {\rm Var}(H_{\rm co}) = A^2 \, {\rm Var}( h_{\rm max}) . \nonumber
\end{eqnarray}
We assumed that sample-to-sample fluctuations solely come from $H_{\rm co}$, which is confirmed numerically.
Since the initial distribution of $h_i$ is Gaussian, $h_{\rm max}$ follows a Gumbel distribution, associated with extreme-value 
statistics~\cite{fortin2015applications}. We then arrive at
\begin{equation}
\chi^{\rm OP}_{\rm dis}(R,L)=N {\rm Var}(\Delta m_{\rm max})= \frac{A^2 \pi^2 R^2 N}{12 \ln N} \ ,
\end{equation}
with $N=L^d$. This expression explains the peculiar growth of the variance with $L$ when $R$ gets smaller in Fig.~\ref{fig:critical_point_delta_m}(c, d).
This effect appears much more pronounced in the $2d$ case.

\begin{figure}
    \centering
\includegraphics[width=0.9\linewidth]{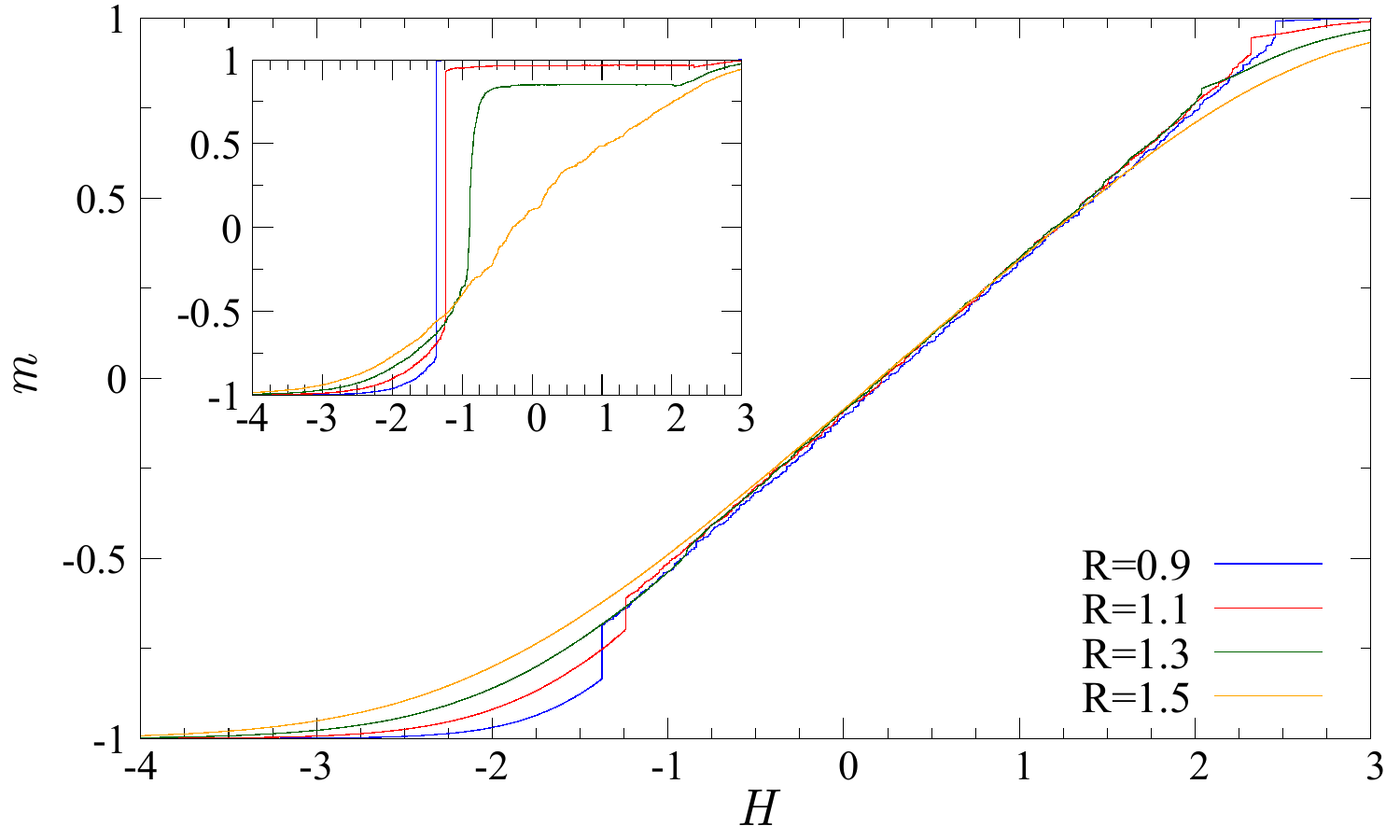}
\caption{Magnetization curves $m(H)$ for $3d$ samples with $L=104$ and different values of $R$. Insets: The corresponding $m_{\rm b}(H)$ curves.}
\label{fig:magnetization_diffR}
\end{figure}

To avoid the above problem, which is an artifact of lattice-based modeling, we introduce an alternative order parameter. As discussed in the main text 
for the description of the discontinuous regime at small disorder, we monitor the mean magnetization along horizontal and vertical lines (in $2d$) or planes 
(in $3d$),  
\begin{align}
    m_x &= \frac{1}{L} \sum_{y=1}^L s_{(x,y)} \qquad (2d), \\ 
    m_x&=\frac{1}{L^2}\sum_{y,z=1}^L s_{(x,y,z)} \qquad (3d),  
\end{align}
and similarly for $m_y$ (and $m_z$ in $3d$). When the disorder strength $R$ is large, spins flip rather homogeneously in space, and no noticeable 
difference is observed between these quantities, {\it e.g.},  in $3d$, between $m_x$, $m_y$, and $m_z$ whatever $x,y,z$. On the other hand, when $R$ 
is small, a sharp band emerges, causing a strong localization and anisotropy of spin flips. In such a situation, there is some location, say $x^*$, for 
which $m_{x^*} \simeq 1$ right after the macroscopic magnetization jump. Thus, we monitor at each step $t$ of the evolution (see Sec.~\ref{app_sub:model}) 
$m_{\rm b,t }= \max_{x,y}\{m_{x, t},m_{y, t}\}$, or its counterpart in $3d$, which precisely detects the emergence of a band at $H_{\rm co}$ (if present). 
Some representative magnetization curves $m(H)$ for various $R$ and the corresponding $m_{\rm b}(H)$ are shown in Fig.~\ref{fig:magnetization_diffR}.

To characterize the disorder-controlled transition between the discontinuous and continuous regimes, and as we did for the magnetization, we then compute 
the maximum jump as $\Delta m_{\rm b, max}=\max_t \{ \Delta m_{\rm b, t }\}$ with $\Delta m_{\rm b, t} = m_{\rm b,t+1}-m_{\rm b,t}$ for each sample realization. 
From this we define the sample-average order parameter $\overline{\Delta m_{\rm b, max}}$ and the associated disconnected susceptibility 
$\chi^{\rm OP}_{\rm dis, b}(R,L)=L^{d-1} {\rm Var}(\Delta m_{\rm b,max})$. 

Note that the disconnected susceptibilities defined from either $m_{\rm b}(H)$ or $\Delta m_{\rm b, max}$ are equal to the variance multiplied by $L^{d-1}$ 
and not by $N=L^d$ as for the magnetization and its maximum jump. The difference is due to the fact that $m$ is a volume-averaged quantity in each sample 
whereas $m_{\rm b}$ is a line- or plane-averaged quantity. In the limit where all spins are uncorrelated, it is easy to see that the so-defined disconnected 
susceptibilities are of O(1), which provides the proper baseline for studying their peak.
\\ 

\subsubsection{Discontinuous regime}

\begin{figure}
\centering
\includegraphics[width=\linewidth]{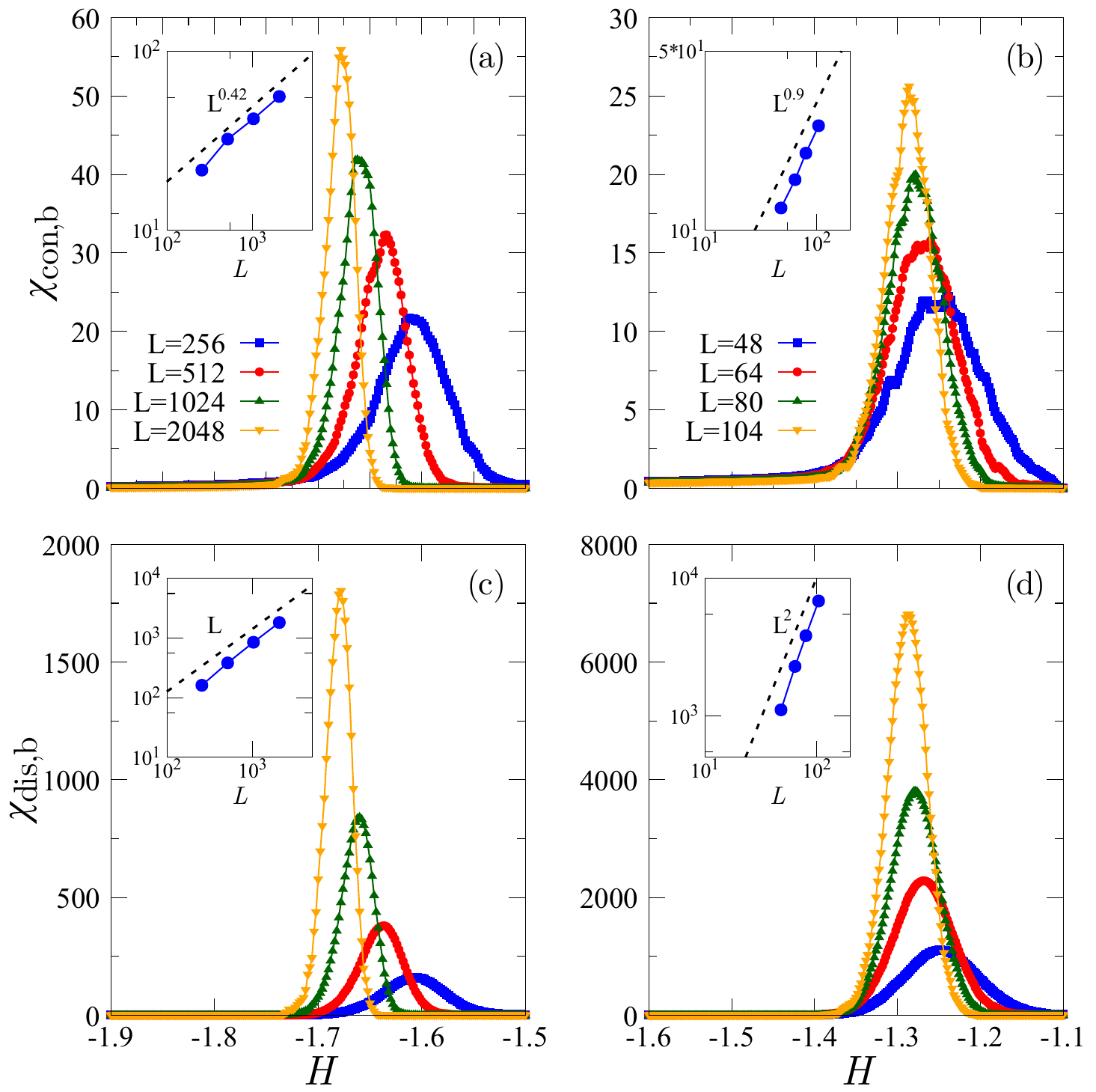}
\caption{Weak-disorder regime: Finite-size scaling analysis of the disconnected and connected susceptibilities  $\chi_{\rm con,b}(H,L)$ (top) and  
$\chi_{\rm dis,b}(H,L)$  (bottom) in $2d$ for $R=0.64$ (left) and in $3d$ for $R=1.05$ (right). Insets: log-log plot of the susceptibility peak versus 
system size $L$. The peak of the disconnected ones grows with an exponent $d-1$ and that of the connected one with an exponent around $0.42$ in $2d$ and 
$0.9$ in $3d$.}
\label{fig:app_discontinuous}
\end{figure}

Before discussing the disorder-controlled transition we first illustrate the existence of a discontinuous regime at small disorder. As mentioned in the main text 
the magnetization discontinuity observed in each sample is rounded when one averages over samples. One then has to proceed to a finite-size scaling analysis 
to check whether the discontinuity reappears in the thermodynamic limit. We compute the disconnected and connected susceptibilities associated with the two 
order parameters $m(H)$ and $m_{\rm b}(H)$,
\begin{equation}
\begin{aligned}
\label{eq:app_susceptibilities}
&\chi_{\rm dis}(H)=L^d {\rm Var}(m(H)),  \\& \chi_{\rm con}(H)=\frac{\partial \overline{m(H)}}{\partial H},  \\&
\chi_{\rm dis, b}(H)=L^{d-1} {\rm Var}(m_{\rm b}(H)),  \\& \chi_{\rm con, b}(H)=\frac{\partial \overline{m_{\rm b}(H)}}{\partial H},
\end{aligned}
\end{equation}
for a small $R$ and for several system sizes $L$. We illustrate in Fig.~\ref{fig:app_discontinuous} the outcome for the susceptibilities associated with 
$m_{\rm b}(H)$ for $R=0.64$ in $2d$ and $R=1.05$ in $3d$. We can see that the suceptibilities peak at a well-defined coercive field 
$H_{\rm co}(L)$ and that the peak sharpens and gets higher as $L$ increases. We find that the peak maximum of $\chi_{\rm dis, b}$ grows as 
$L^{d-1}$ both in $2d$ and $3d$ while the maximum of $\chi_{\rm con, b}$ grows with $L$ with smaller exponents due to the sample-to-sample fluctuations of 
$H_{\rm co}$~\cite{ozawa2020role}: see the insets. A similar behavior is found for the susceptibilities associated with $m(H)$ except that the peak of 
$\chi_{\rm dis}(H)$ grows as $L^d$.

\subsubsection{Disorder-controlled transition}

To characterize the critical behavior (exponents and scaling functions) from a finite-size scaling analysis, we have used several methods based on 
the scaling ansatz described in the main text [see the relations in Eq.~(2)]. Note that extracting the critical properties from 
finite-size studies is much more demanding (and as a result less accurate) for a far-from-equilibrium critical point in a disordered system than for 
conventional critical points in pure systems at equilibrium. It indeed lacks the symmetries (time-reversal, inversion) of the equilibrium ones and is 
affected by strong corrections to scaling. This has been vividly illustrated in the case of scaling analyses of the mean-field driven RFIM, whose 
analytical solution is otherwise exactly known~\cite{perkovic1996,*perkovic1999}.

\begin{figure}
\centering
\includegraphics[width=\linewidth]{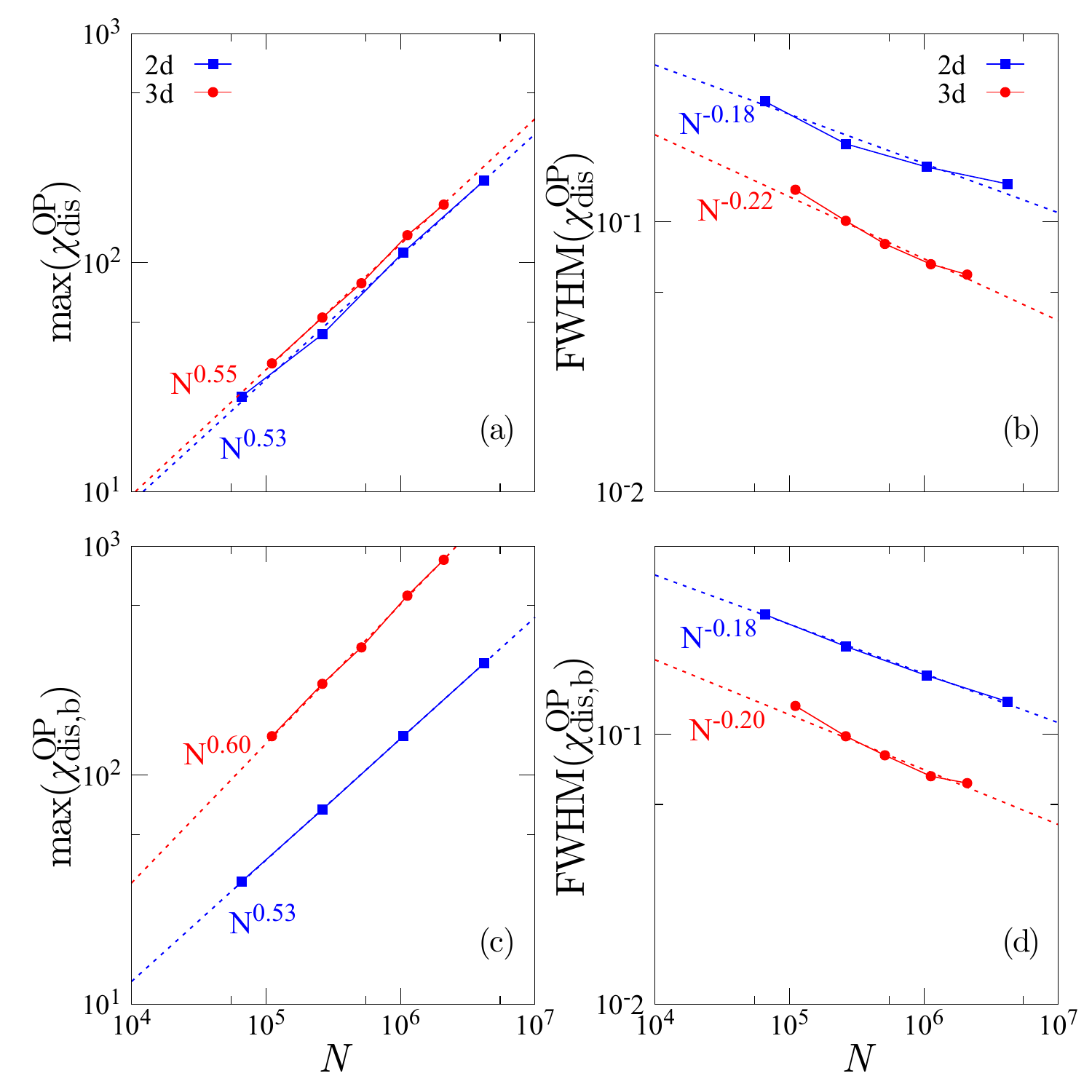}
\caption{Log-log plot of the height (left) and width (right) of the $R$-dependent disconnected susceptibilities $\chi_{\rm dis}^{\rm OP}(R, L)$ (top) and 
$\chi_{\rm dis,b}^{\rm OP}(R, L)$ (bottom) as a function of the system size $N=L^d$ in $2d$ and $3d$.}
\label{fig:app_log-log_OP}
\end{figure}

\begin{figure}
\centering
\includegraphics[width=0.6\linewidth]{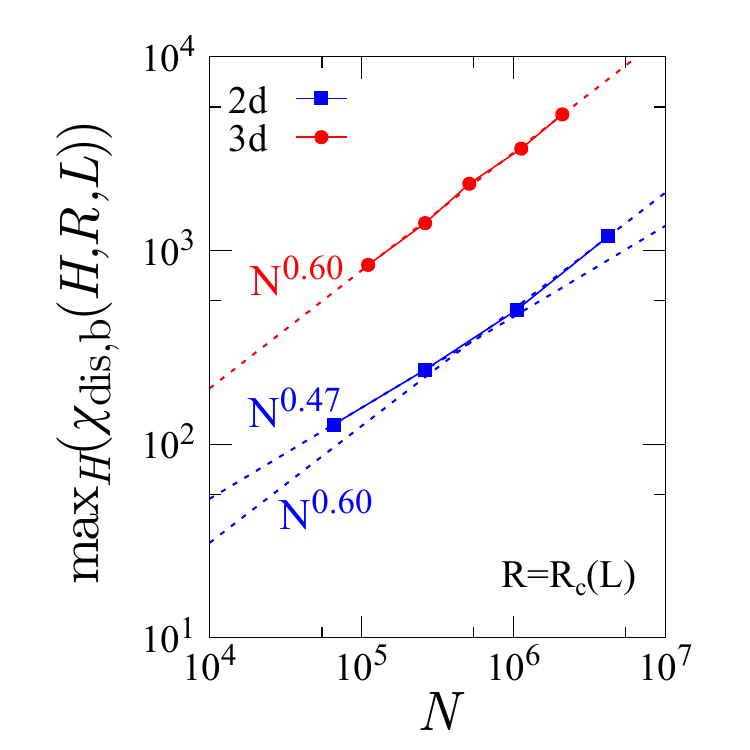}
\caption{Log-log plot of the height of the maximum over $H$ of the disconnected susceptibility $\chi_{\rm dis,b}(H,R, L)$ at $R=R_c(L)$ as 
a function of $N=L^d$ in $2d$ and $3d$.
}
\label{fig:app_log-log}
\end{figure}

The first method we used is to consider the height and the width of the $R$-dependent disconnected susceptibilities $\chi_{\rm dis}^{\rm OP}(R, L)$ and 
$\chi_{\rm dis,b}^{\rm OP}(R, L)$ as a function of the system size: The height is expected to grow as $L^{\bar\gamma/\nu}$ or $L^{\bar\gamma_{\rm b}/\nu}$ 
and the width to decay as $L^{-1/\nu}$. We show a log-log plot of the dependence on $N=L^d$ for both $2d$ and $3d$ in Fig.~\ref{fig:app_log-log_OP}. 
This allows us to estimate the critical exponents as: $\bar \gamma/\nu\approx 1.07 \pm 0.03$, $\bar \gamma_{\rm b}/\nu\approx 1.060\pm 0.002$, 
$\nu\approx 2.74\pm0.05$ in $2d$ and $\bar\gamma/\nu\approx 1.65\pm0.06$, $\bar \gamma_{\rm b}/\nu\approx 1.82\pm0.05$, and $\nu\approx 1.64\pm0.06$
in $3d$, where the error bars are derived from the fits. We have also checked that a similar procedure applied to the maximum over $H$ of the disconnected 
susceptibility $\chi_{\rm dis,b}(H, L)$  evaluated for $R=R_c(L)$ gives compatible results ($\bar\gamma_{\rm b}/\nu $ between 0$.95$ and $1.2$  in $2d$ and $\bar\gamma_{\rm b}/\nu \approx 1.80 \pm 0.03$ in $3d$), as illustrated in 
Fig.~\ref{fig:app_log-log}.

\begin{figure}
\centering
\includegraphics[width=\linewidth]{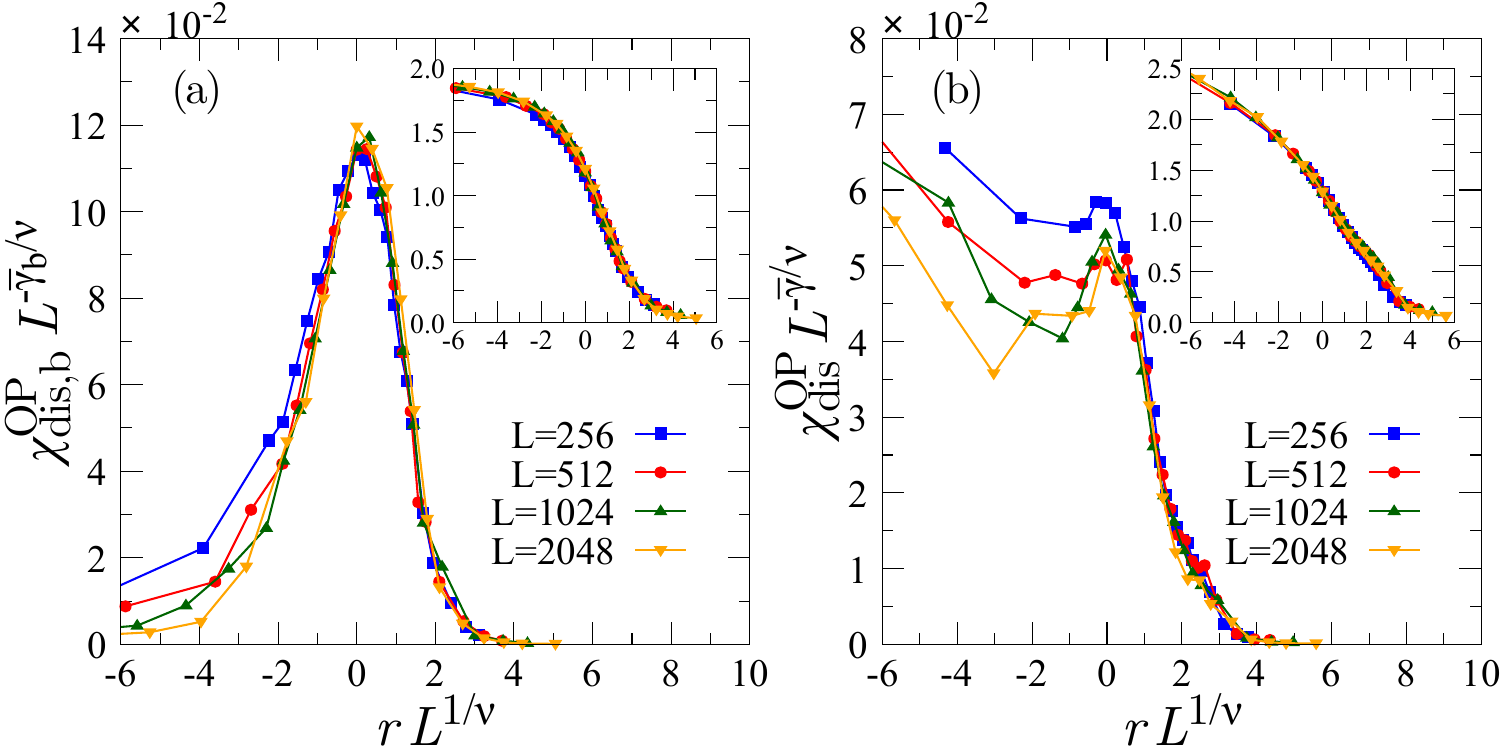}
\caption{Finite-size scaling collapse of the disconnected susceptibilities $\chi^{\rm OP}_{\rm dis,b}(r,L)$  with $\bar\gamma_{\rm b} / \nu= 1.03$ (a) 
and $\chi^{\rm OP}_{\rm dis}(r,L)$ with $\bar\gamma/\nu=1.1$ (b) in $2d$, where $r=(R-R_c(L))/R$. 
Insets: Same for $\overline{\Delta m_{\rm b,max}}$ with $\beta_{\rm b}/\nu=0.0$ (a) and $\overline{\Delta m_{\rm max}}$ with 
$\beta/\nu=0.46$ (b). In all cases $\nu=2.7$.}
\label{fig:app_collapse_2d}
\end{figure}

We next perform scaling collapses to the expressions in Eq.~(2) of the main text, where the exponents and the critical disorder $R_c(L)$ are adjusted to 
provide the best visual collapse of the curves obtained for different values of $L$. This gives values that are consistent with the previous ones, 
$\bar\gamma/\nu\approx 1.1\pm0.05$, $\bar \gamma_{\rm b}/\nu\approx 1.03\pm0.03$, $\nu\approx 2.7\pm0.1$ in $2d$ and 
$\bar\gamma/\nu\approx 1.70\pm0.05$, $\bar \gamma_{\rm b}/\nu\approx 1.81\pm0.05$, and $\nu\approx 1.8\pm0.6$ in $3d$. In addition, from the collapse 
of the disorder-averaged order parameters, we also obtain $\beta/\nu\approx 0.46\pm0.02$, $\beta_{\rm b}/\nu\approx 0.0\pm0.02$, $\nu\approx 2.7\pm0.2$ 
in $2d$, and $\beta/\nu\approx 0.62\pm0.04$, $\beta_{\rm b}/\nu\approx 0.08\pm0.04$, and $\nu\approx 1.4\pm0.4$ in $3d$. The results are shown for $3d$ 
in the main text and we give here their $2d$ counterparts: see Fig.~\ref{fig:app_collapse_2d}.

We also fit the disconnected susceptibility data to a flexible functional form of master-curve to extract the universal scaling function. To start with, we use a 
nonlinear fit with a Gaussian around the peak of the susceptibility curves to get an estimate of the value of $R_c(L)$. We find that the estimated value is very 
close to the one obtained by simply looking at the peak position (see above). Following the same procedure as in Ref.~[\onlinecite{sethna2013percolationfailure}] we 
fit the curves for $\chi_{\rm dis,b}^{\rm OP}(r,L)$ with $A_{\rm b} L^{\bar\gamma_{\rm b} / \nu}\bar\Psi_{\rm b}(\alpha_{\rm b} r L^{1 / \nu})$, where 
we choose
\begin{equation}
\label{eq:scaling_form}
   \bar \Psi_{\rm b}(x) = \exp{( x-\exp{( x)})}\left(1+\sum_{i=1}^{3} a_{{\rm b},i} H_i(x)\right),
\end{equation}
with $H_i(x)$ the $i$-th Hermite polynomial  and $\bar\gamma_{\rm b}/\nu$, $\nu$, $A_{\rm b}$, $\alpha_{\rm b}$, and $a_{{\rm b},i}$ ($i=1,2,3$) are adjustable 
parameters. $A_{\rm b}$ and $\alpha_{\rm b}$ control the axis scales and are irrelevant for the universality class, while the $a_{{\rm b},i}$'s describe the curve 
shape and should be universal. To reduce the effect of the scaling corrections at smaller sizes we only fit systems with $L \geq 1024$ in $2d$ and $L \geq 80$ 
in $3d$. We then find $\bar\gamma_{\rm b}/ \nu = 1.07 \pm 0.04$, $\nu = 2.7 \pm 0.4$ for $2d$ and $\bar\gamma_{\rm b}/ \nu = 1.87 \pm 0.03$ and $\nu = 1.6 \pm 0.4$ 
for $3d$.

Finally, we have considered the shift of the critical disorder and fitted it to $R_c(\infty)-R_c(L)\propto L^{-1/\nu}$. This is described in Sec.~\ref{app:bounding}.

\subsubsection{Comparison with elasto-plastic models}

To investigate whether the critical points of the RFIM-Eshelby and of the mesoscopic models for sheared amorphous solids known as elasto-plastic models 
(EPMs) are in the same universality class we have compared the scaling functions for the disconnected susceptibilities.

\begin{figure}
    \centering
\includegraphics[width=\linewidth]{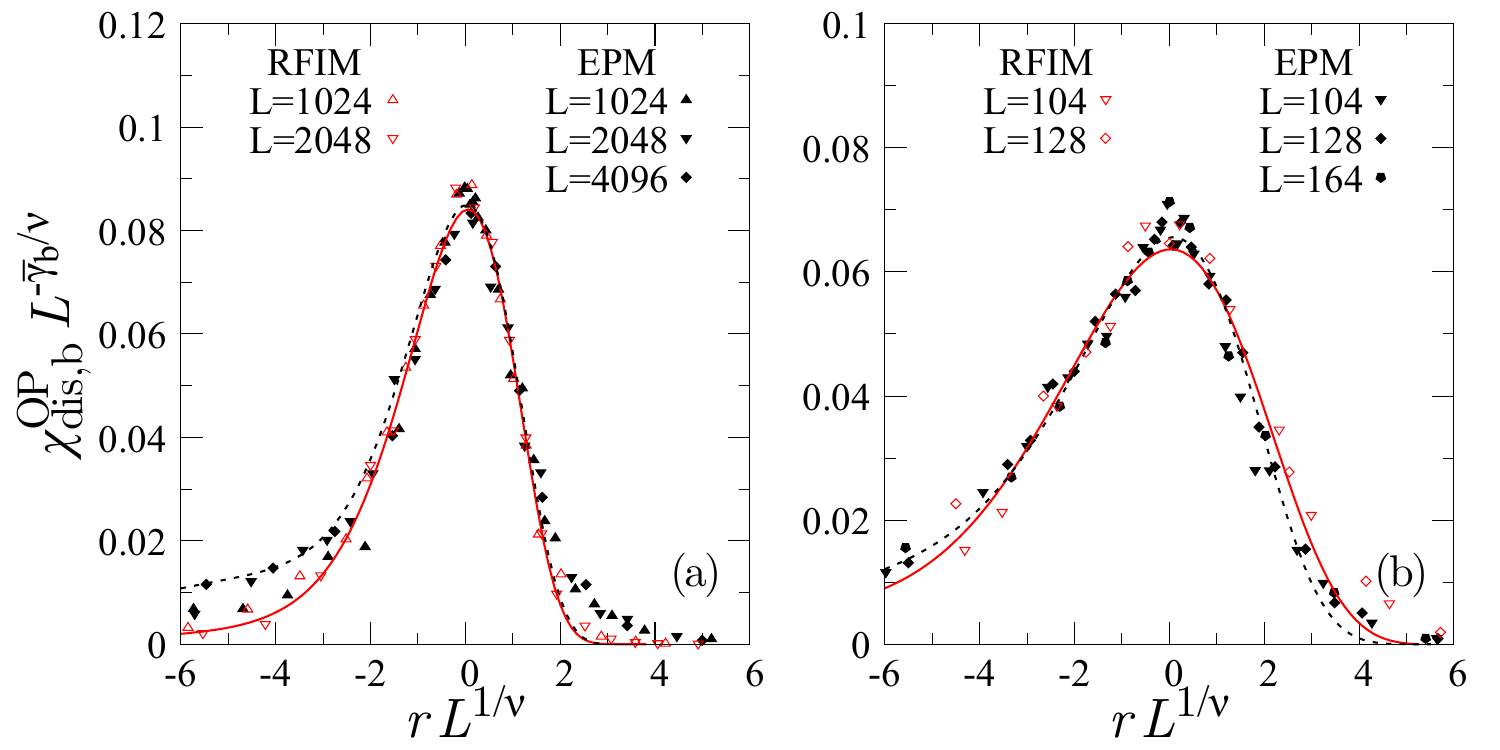}
\caption{Comparison of the scaling functions for the disconnected susceptibility $\chi_{\rm dis,b}^{\rm OP}(r,L)$ of the Eshelby-RFIM (red) and of the 
elasto-plastic model of Ref.~[\onlinecite{rossi2022finite}] (black) in $2d$ (left) and $3d$ (right). The full red (dashed black) line is the fitted functional form of Eq.~(\ref{eq:scaling_form}) to the data coming from the Eshelby-RFIM (EPM), 
and the data are only shown for $L\ge 1024$ in $2d$ and $L \ge 104$ in $3d$. The ratio $\bar\gamma_{\rm b}/\nu$ used for the collapse is $10\%$ higher in the Eshelby-RFIM.
}
\label{fig:scaling_functions}
\end{figure}

We first proceed to the same treatment as above with the parametrization in Eq.~(\ref{eq:scaling_form}) to extract the scaling function of 
$\chi_{\rm dis,b}^{\rm OP}(R,L)$ (obtained from the variance of the maximum jump of the band order parameter $n_{\rm b}$ which is the counterpart of $m_{\rm b}$) 
for the EPM studied in Ref.~[\onlinecite{rossi2022finite}]. In this case we fit the curves in $2d$ for $L$ from $1024$ to $4096$ and in $3d$ for $L$ from $80$ to $164$. We plot the result for the scaling functions of the 
Eshelby-RFIM and the EPM in Figure~\ref{fig:scaling_functions}. We see that the two models seem to have a very similar scaling function, but as discussed in 
the main text the exponent ratio $\bar\gamma_{\rm b}/ \nu$ is about $10\%$ more in the RFIM.

\begin{figure}
    \centering
\includegraphics[width=\linewidth]{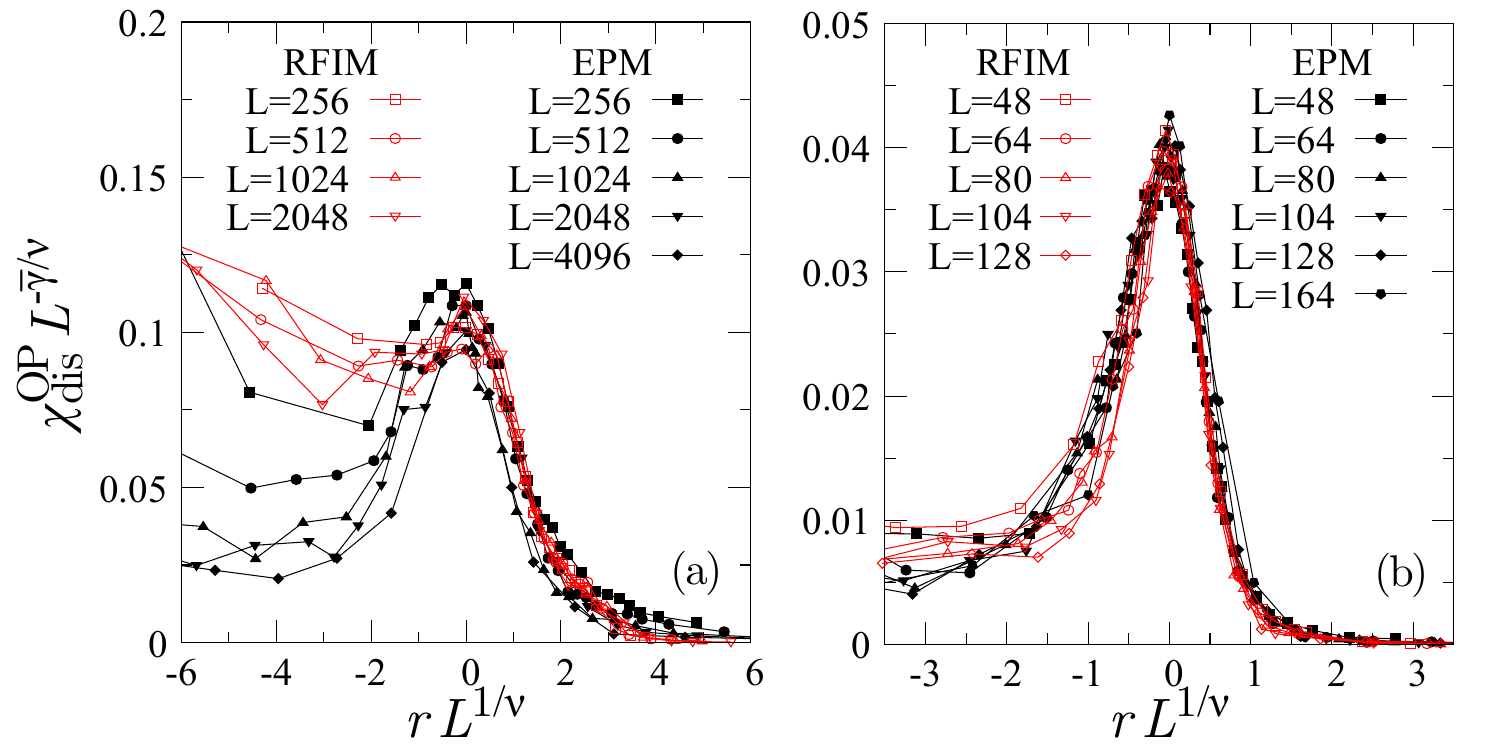}
\caption{Finite-size scaling collapse of the disconnected susceptibility curves $\chi_{\rm dis}^{\rm OP}(r,L)$ for the Eshelby-RFIM (empty symbols, red) and the 
elasto-plastic model  of Ref.~[\onlinecite{rossi2022finite}] (full symbols, black) in $2d$ (left) and $3d$ (right). The collapse is obtained by using the same exponents for the two 
models: $\bar\gamma/ \nu=1.0$ and $\nu=2.7$ in $2d$, $\bar\gamma/ \nu=1.75$ and $\nu=1.6$ in $3d$. }
\label{fig:scaling_collapse}
\end{figure}

We also perform a direct comparison of the disconnected susceptibility curves $\chi_{\rm dis}^{\rm OP}(r,L)$ built from the maximum 
magnetization jump (RFIM) or from the maximum stress drop (EPM) for $2d$ and $3d$. We collapse all the curves by using the {\it same} exponents for the 
Eshelby-RFIM and the EPM, namely, $\bar\gamma/ \nu=1.05$ and $\nu=2.7$ in $2d$ and $\bar\gamma/ \nu=1.75$ and $\nu=1.6$ in $3d$. As seen in Fig.~\ref{fig:scaling_collapse}, the collapse appears very good in $3d$ (less so in $2d$ where the curves are significantly affected by the extreme-value 
statistics at low disorder). These data also suggest that the two models are in the same universality class.

\subsection{Bounding the location of the disorder-controlled critical point}
\label{app:bounding}

\begin{figure}
    \centering
\includegraphics[width=\linewidth]{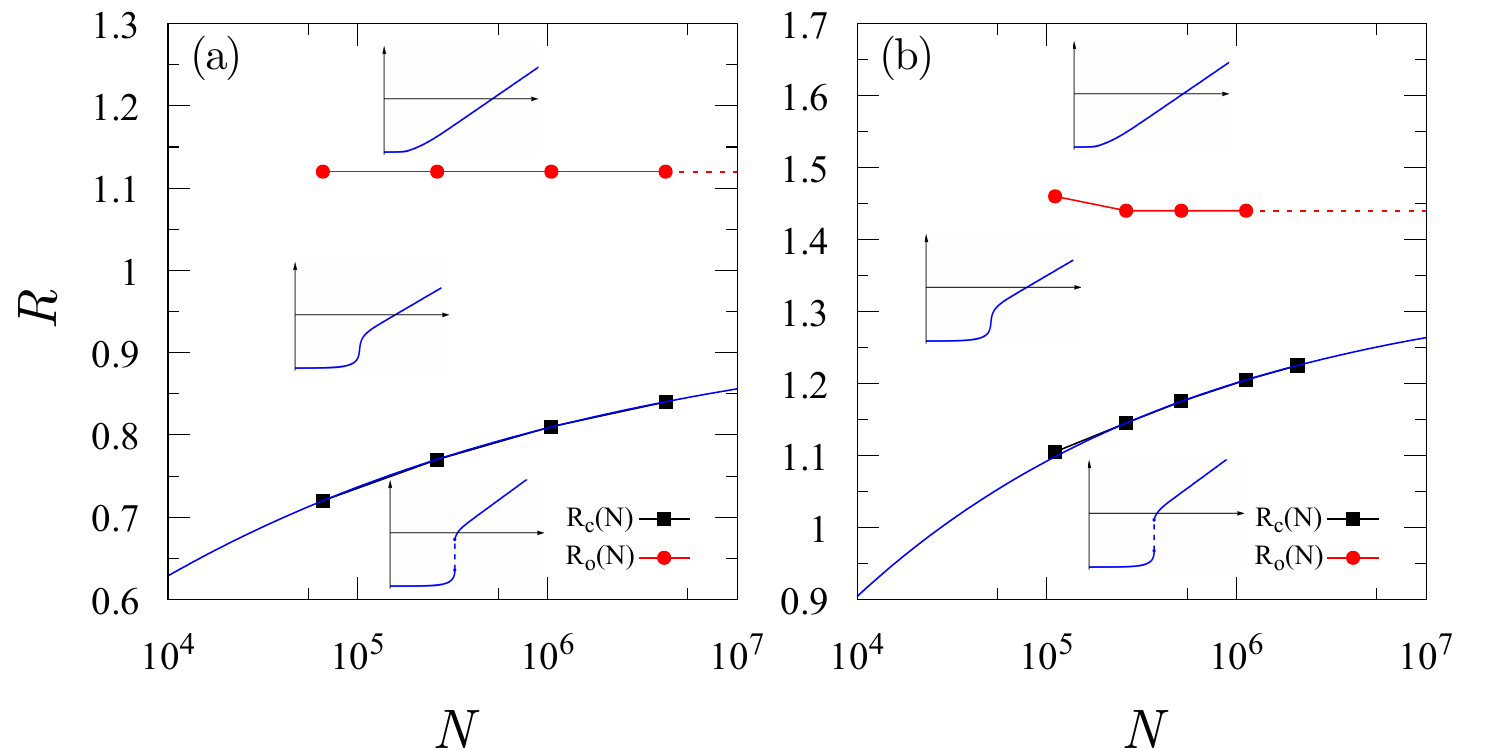}
\caption{The critical point, $R_c(N)$, as a function of the system size, $N=L^d$, together with the value of the disorder $R_o(N)$ at which an overshoot first 
appears in the transformed magnetization curve $\tilde\sigma(H,N)$ for the $2d$ (a) and the $3d$ (b) cases. Blue curves are fits to $R_c(\infty) - R_c(N) = aN^{-b}$, 
with $a$, $b$, and $R_c(\infty)$ taken as adjustable parameters and given in the text.
Insets: The corresponding schematic magnetization curves.} 
\label{fig:Rc_vs_Ro}
\end{figure}

Figure~\ref{fig:Rc_vs_Ro} presents $R_c(N)$ as a function of $N=L^d$ in both $2d$ and $3d$, showing a slow monotonic increase of $R_c(N)$ 
with $N$. We now provide evidence that $R_c$ remains finite in the thermodynamic limit. Note that the fate of the critical point in the thermodynamic 
limit has been a key issue in the study of yielding in amorphous materials~\cite{barlow2020ductile,richard2021finite,rossi2022finite}.
As discussed above, a standard finite-size scaling argument suggests that  $R_c(\infty) - R_c(L) \propto L^{-1/\nu}$. Through a direct fit of 
$R_c(\infty) - R_c(N) = a N^{-b}$, with $a$, $b$ and $R_c({\infty})$ adjustable parameters, we find $R_c({\infty}) = 0.95$, $a=1.65$, $b=0.178$ in 
$2d$ and $R_c({\infty}) = 1.35$, $a=3.97$, $b=0.237$ in $3d$. The parameter $b$ is related to the critical exponent $\nu$ through $b=1/(d\nu)$, which 
yields $\nu \approx 2.80$ in $2d$ and $\nu \approx 1.41$ in $3d$, values which are compatible with the independent finite-size analysis performed above. 

\begin{figure}
    \centering
\includegraphics[width=\linewidth]{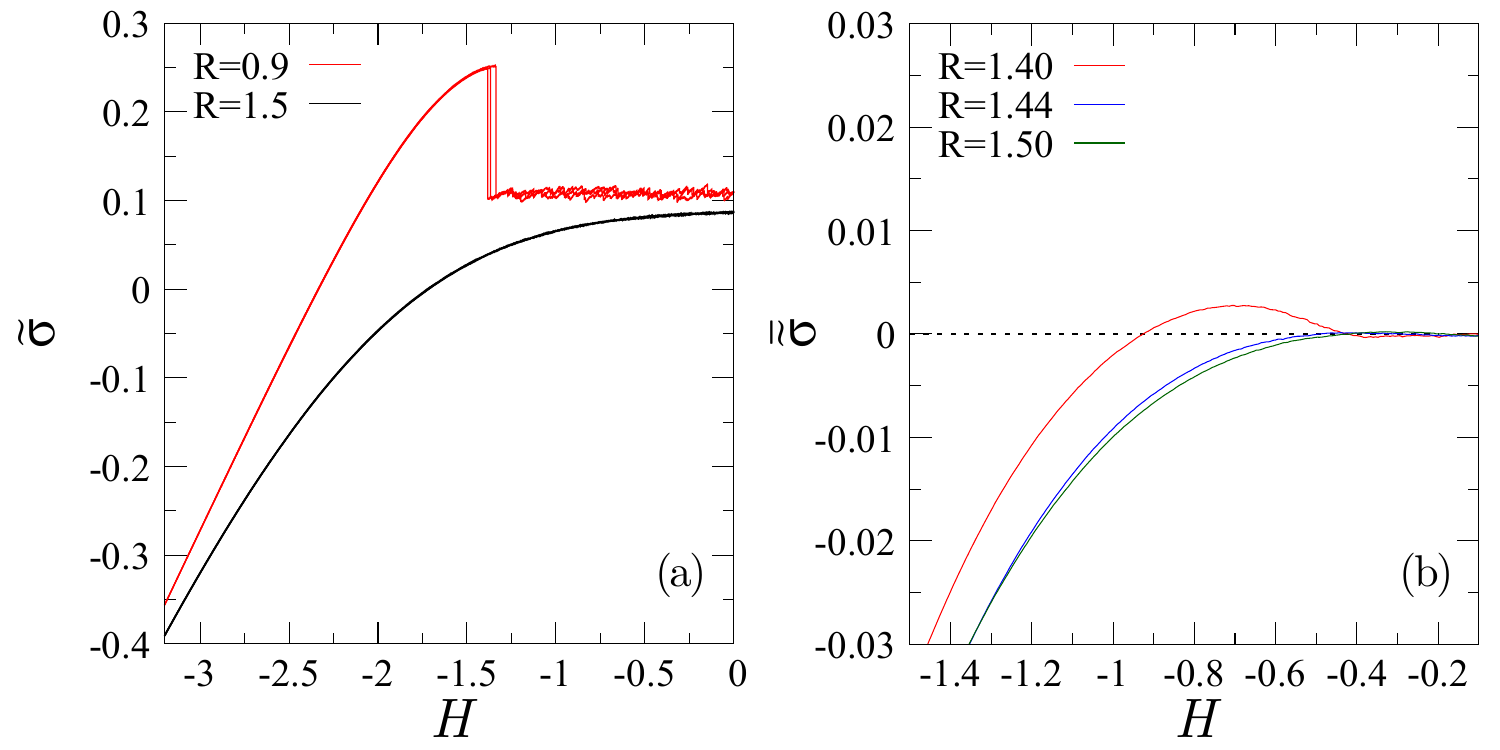}
\caption{(a): The transformed magnetization curve, $\tilde\sigma(H)=A H-m(H)$, for the $3d$ Eshelby-RFIM with a small and a large disorder strength $R$ (same samples as in Fig.~1(a) of the main text). Note the similarity with the stress-strain curves of a mesoscopic model of a uniformly sheared amorphous solid~\cite{rossi2022finite}. 
(b): The onset value $R_o(L)$ at which a local maximum (overshoot) first appears is identified at $R_o\approx1.5$ for $L=104$.}
\label{fig:Ro_determination}
\end{figure}

To go beyond this fit and more firmly establish that $R_c(\infty)$ has a finite value, we estimate an upper bound following the method devised in 
the study of yielding in an EPM~\cite{rossi2022finite}. In the latter we monitored the value $R_o(N)$ at which an overshoot, {\it i.e.}, a local maximum, first appears 
in the disorder-averaged stress ($\sigma$) versus strain ($\gamma$) curve. This value is always an upper bound of the critical disorder. We found that $R_o(N)$ is 
independent of $N$ over the whole accessible range of system sizes both in $2d$ and $3d$~\cite{rossi2022finite}. To repeat this analysis in the case of the Eshelby-RFIM we first 
build on the analogy between the two models already described in the main text. As illustrated in Fig. 1 of the main text and here in 
Fig.~\ref{fig:magnetization_diffR}, the magnetization follows a linear regime $m(H)=A H+B$ (with $A$ and $B$ weakly $R$-dependent) before saturating to 
$+1$ at large applied field $H$. When plotting the quantity $\tilde\sigma(H)=A H-m(H)$ we can see that it qualitatively reproduces the stress-strain curve 
$\sigma(\gamma)$ of the EPM with the same evolution as a function of disorder strength: see Fig.~\ref{fig:Ro_determination}(a) for an illustration in $3d$. From 
the average over samples of these transformed magnetization curves we then identify the onset $R_o(N)$ at which a local maximum (overshoot) first appears in $\tilde\sigma(H,R)$, as shown 
in Fig.~\ref{fig:Ro_determination}(b). The outcome is plotted in Fig.~\ref{fig:Rc_vs_Ro} and we find that, as for the EPM,  $R_o(N)$ is essentially independent of 
$N$ and therefore provides a finite upper bound for the critical disorder $R_c(\infty)$.
\\

\subsection{Avalanches at small disorder: a comparison between the Eshelby-RFIM and the EPM}
\label{app:Avalanches}

The evolution of AQS driven disordered systems proceeds by avalanches. We focus here on a comparative study of the Eshelby RFIM and the EPM of 
Ref.~[\onlinecite{rossi2022finite}] in the small-disorder regime, away from the disorder-controlled critical point. The EPM is considered at large enough imposed strain, beyond the yielding transition and the Eshelby-RFIM in the linear regime beyond the coercive field $H_{\rm co}$ and well before the saturation 
to a magnetization $m=+1$.

We study the distribution of avalanche sizes $P(S)$ for a given (small) $R$ and cumulated along a significant fraction of the linear regime between two values, 
$m_{\rm min}$ and $m_{\rm max}$, of the average magnetization. The size $S$ of an avalanche is defined as $N$ times the associated jump $\delta m$ of the 
sample-dependent magnetization and $P(S)$ is normalized. In AQS driven disordered systems one commonly encounters a scaling behavior of the form 
\begin{equation}
\label{eq:appD_scalinglawavalanches}
P(S) \sim S^{-\tau} \mathcal{P}(S/S_c),
\end{equation} 
with $\tau$ an exponent and $\mathcal{P}$ a scaling function that decreases very quickly for $S>S_c$. In a finite system of linear size $L$, the cutoff avalanche 
size $S_c$ can either be independent of $L$ or, in the presence of criticality (either disorder-controlled, self-organized, or due to marginal stability), grow as 
$S_c \sim L^{d_f}$, where $d_f$ is the fractal dimension of the largest avalanches. This can be probed by computing $P(S)$ and trying to collapse all the 
curves on a plot of $P(S)L^{\tau d_f}$ vs $S/L^{d_f}$ by adjusting $\tau$ and $d_f$. These exponents have been extensively studied for various AQS driven 
disordered solids such as the ferromagnetic RFIM at criticality~\cite{perkovic1996,*perkovic1999,vives-perez03,vives-perez04}, elastic manifolds at their depinning 
transition (for a review, see~[\onlinecite{wiese22}]), or sheared amorphous solids~\cite{Salerno2013MDresultsforavalanches,Lin2014dynamicalyieldingtransitionEPM,nicolas2018deformation}. 

In the EPM description of sheared 
amorphous solids in their steady state, these scale-free avalanches are associated with a marginal stability according to which there exists a pseudo-gap 
in the distribution of the local stability, $\tilde P(x)\sim x^\theta$ with $\theta>0$~\cite{lerner-procaccia09,Lin2014dynamicalyieldingtransitionEPM}. This property 
also translates in the distribution of the strain intervals between two successive avalanches $Q(\delta\gamma)$, for which the corresponding average interval 
goes to zero with system size as $<\delta\gamma>\sim N^{-1/(1+\theta)}$. In $2d$ EPMs one finds $\theta\approx 0.5-0.6$ and in $3d$ 
$\theta\approx 0.3-0.45$~\cite{lerner-procaccia09,Lin2014dynamicalyieldingtransitionEPM,Lin2014theta,nicolas2018deformation}. 

\begin{figure}[!t]
\centering
\includegraphics[width=0.95\linewidth]{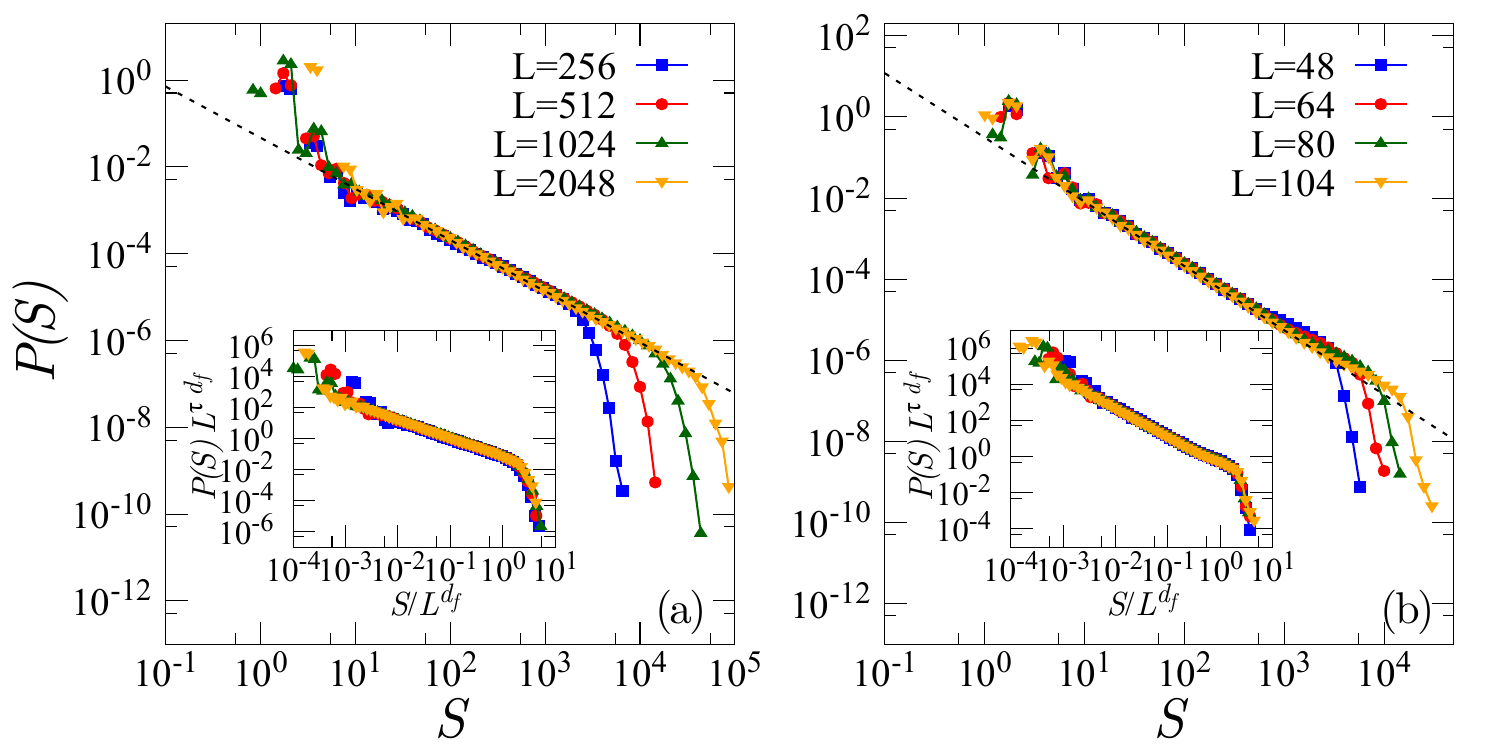}
\caption{Finite-size scaling of the avalanche size distribution for $2d$ and $R=0.6$ (a) and $3d$ and R=0.8 (b). The vertical axis is shifted so that the large 
avalanches ($S\gtrsim20$) fall as much as possible on the same curve, which is marked by a black dashed line. The insets show a further collapse by rescaling 
the axes. The collapse gives $\tau \approx 1.22$,  $d_f\approx1.3$ in $2d$ and $\tau \approx 1.58$,  $d_f\approx1.9$ in $3d$.}
\label{fig:app_interval-scaling_RFIM.}
\label{fig:app_avalanche-scaling_RFIM}
\end{figure}

\begin{figure}[!t]
\centering
\includegraphics[width=0.95\linewidth]{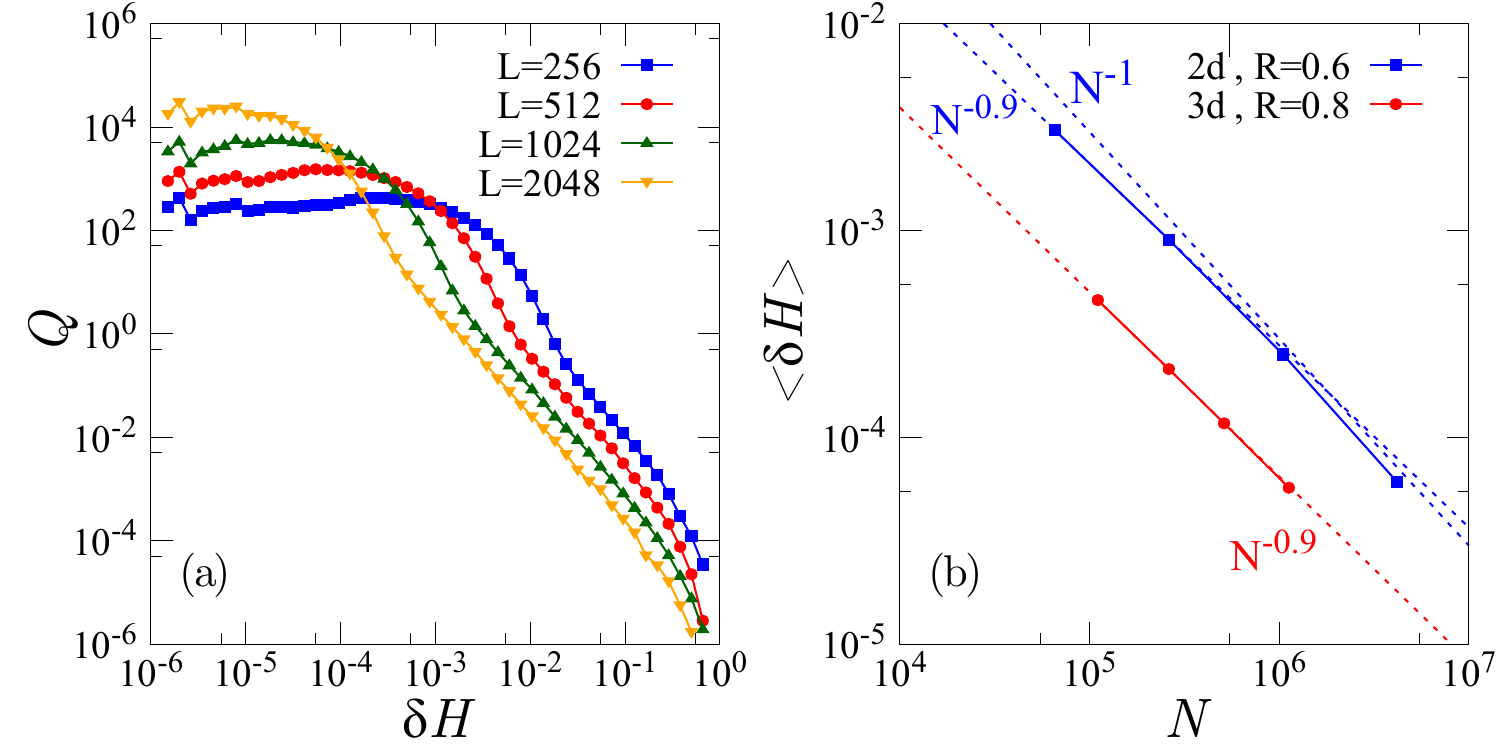}
\caption{(a): Distribution $Q(\delta H,L)$ of the intervals between successive avalanches in the linear regime of the $2d$ Eshelby-RFIM with $R=0.6$. 
(b): Log-log plot of the average interval $<\delta H>$ versus $N$ for the same case and for the $3d$ version at $R=0.8$. The dashed lines indicate $N^{-1}$ and $N^{-0.9}$ for $2d$ and $N^{-0.9}$ for $3d$.}
\label{fig:app_interval-scaling_RFIM}
\end{figure} 

We have tested our procedure to determine the avalanche size distribution $P(S)$ and the distribution of intervals between successive avalanches 
$Q(\delta\gamma)$ for the EPM already studied in Ref.~[\onlinecite{rossi2022finite}] and we have properly recovered the results of the existing literature. We 
have then applied the same treatment to the Eshelby-RFIM in the linear regime of the magnetization curve. A difficulty, however, is that there seems 
to be a different behavior of the small (say, $S\lesssim 20$) and the large ($S\gtrsim 20$) avalanches as the system size increases. One possible explanation is that because the interactions are not purely ferromagnetic in the Eshelby-RFIM, a spin can flip more 
than once. In the case of a $1d$ long-ranged anti-ferromagnetic RFIM, this was shown to lead to different types of avalanches with different characteristics\cite{chakraborty17}. However, in the present case we find that spins flip usually once and at most a few number 
of times. To nonetheless see if there is a difference between avalanches involving a change of magnetization and avalanches 
in which spin flips do not change the magnetization we have considered a measure of the avalanche size by the number of spin flips, but we have 
obtained essentially the same result as with the magnetization jumps. One can try to somehow overlook the small avalanches (despite the fact that their 
number grows with the system size) by collapsing the avalanche distribution only in the range of the large avalanches: this is shown in 
Fig.~\ref{fig:app_avalanche-scaling_RFIM} for weak disorder in $2d$ and $3d$. One can fit a power law over some range with exponent $\tau\approx1.2$ 
in $2d$ and $1.6$ in $3d$, and the cutoff avalanche size seems to grow with $L$.

We have also monitored the distribution of intervals between avalanches, $Q(\delta H,L)$, and computed the average $<\delta H>$. The distribution is 
displayed in Fig.~\ref{fig:app_interval-scaling_RFIM}(a) and the average interval versus system size is shown on a log-log plot in 
Fig.~\ref{fig:app_interval-scaling_RFIM}(b). We find that $<\delta H>$ varies very weakly with $L$, which would predict an exponent $\theta\sim 0-0.1$, 
barely compatible with any marginal stability. (In $3d$, we find $\theta\approx 0.1$.)

Although a more exhaustive investigation taking into account the properties of the different types of avalanches and of intervals $\delta H$ would be required, 
our results seem to indicate that EPMs and RFIMs away from the disorder-controlled critical point display different types of behavior as far as avalanches, 
marginal stability, and of course the existence or not of a {\it bona fide} steady state, are concerned. This appears likely due to the possibility or not for sites to 
be active an arbitrary large number of times.

\end{document}